\documentclass[aps,prd,twocolumn,floats,amsmath,tightenlines,nofootinbib,
preprintnumbers,superscriptaddress,sort&compress]{revtex4}
\usepackage{hypernat}   
\usepackage{amssymb}
\usepackage{bm}
\usepackage{graphicx}

\catcode`@=11
\let\savesort=\NAT@sort@cites
\newcommand\nosort[1]{\edef\NAT@cite@list{#1}}
\def\citenosort#1{\let\NAT@sort@cites=\nosort \cite{#1}%
   \let\NAT@sort@cites=\savesort}
\catcode`@=12 

\newcommand{\beq}{\begin{equation}}
\newcommand{\eeq}{\end{equation}}
\newcommand{\nn}{\nonumber}
\newcommand{\comment}[1]{}
\newcommand{\N}{{\mathcal N}}

\begin{document}

\preprint{MIT-CTP-3975\qquad  SU-ITP-08/20}

\title{Boltzmann brains and the scale-factor 
cutoff measure of the multiverse}

\author{Andrea De Simone}
\affiliation{Center for Theoretical Physics, Laboratory for Nuclear 
Science, and Department of Physics, \\ Massachusetts Institute of
Technology, Cambridge, MA 02139, USA}

\author{Alan H.~Guth}
\affiliation{Center for Theoretical Physics, Laboratory for Nuclear 
Science, and Department of Physics, \\ Massachusetts Institute of
Technology, Cambridge, MA 02139, USA}

\author{Andrei~Linde}
\affiliation{Department of Physics, Stanford University, Stanford, 
CA 94305, USA}
\affiliation{Yukawa Institute of Theoretical Physics, Kyoto University,
Kyoto, Japan}

\author{Mahdiyar~Noorbala}
\affiliation{Department of Physics, Stanford University, Stanford, 
CA 94305, USA}

\author{Michael P.~Salem}
\affiliation{Institute of Cosmology, Department of Physics and Astronomy, \\
Tufts University, Medford, MA 02155, USA}

\author{Alexander Vilenkin}
\affiliation{Institute of Cosmology, Department of Physics and Astronomy, \\
Tufts University, Medford, MA 02155, USA}

\begin{abstract}
To make predictions for an eternally inflating ``multiverse,''
one must adopt a procedure for regulating its divergent spacetime
volume.  Recently, a new test of such spacetime measures has
emerged: normal observers --- who evolve in pocket universes
cooling from hot big bang conditions --- must not be vastly
outnumbered by ``Boltzmann brains'' --- freak observers that pop
in and out of existence as a result of rare quantum fluctuations. 
If the Boltzmann brains prevail, then a randomly chosen observer
would be overwhelmingly likely to be surrounded by an empty
world, where all but vacuum energy has redshifted away, rather
than the rich structure that we observe. Using the scale-factor
cutoff measure, we calculate the ratio of Boltzmann brains to
normal observers.  We find the ratio to be finite, and give an
expression for it in terms of Boltzmann brain nucleation rates
and vacuum decay rates.  We discuss the conditions that these
rates must obey for the ratio to be acceptable, and we discuss
estimates of the rates under a variety of assumptions.
\end{abstract}

\maketitle


\section{Introduction}

The simplest interpretation of the observed accelerating
expansion of the universe is that it is driven by a constant
vacuum energy density $\rho_\Lambda$, which is about three times
greater than the present density of nonrelativistic matter. 
While ordinary matter becomes more dilute as the universe
expands, the vacuum energy density remains the same, and in
another ten billion years or so the universe will be completely
dominated by vacuum energy. The subsequent evolution of the
universe is accurately described as de Sitter space.

It was shown by Gibbons and Hawking~\cite{GH} that an observer in
de Sitter space would detect thermal radiation with a
characteristic temperature $T_{\rm dS}=H_\Lambda/2\pi$, where
\beq 
H_\Lambda=\sqrt{{8 \pi \over 3} G \rho_\Lambda}
\eeq 
is the de Sitter Hubble expansion rate. For the observed value of
$\rho_\Lambda$, the de Sitter temperature is extremely low,
$T_{\rm dS}= 2.3\times 10^{-30}$~K.  Nevertheless, complex
structures will occasionally emerge from the vacuum as quantum
fluctuations, at a small but nonzero rate per unit spacetime
volume. An intelligent observer, like a human, could be one such
structure. Or, short of a complete observer, a disembodied brain
may fluctuate into existence, with a pattern of neuron firings
creating a perception of being on Earth and, for example,
observing the cosmic microwave background radiation.  Such freak
observers are collectively referred to as ``Boltzmann
brains''~\cite{Rees1,Albrecht}. Of course, the nucleation rate
$\Gamma_{\rm BB}$ of Boltzmann brains is extremely small, its
magnitude depending on how one defines a Boltzmann brain.  The
important point, however, is that $\Gamma_{\rm BB}$ is always
nonzero.

De Sitter space is eternal to the future.  Thus, if the
accelerating expansion of the universe is truly driven by the
energy density of a stable vacuum state, then Boltzmann brains
will eventually outnumber normal observers, no matter how small
the value of $\Gamma_{\rm
BB}$~\citenosort{Susskind,Page1,Page2,Page06,BF06} might be.

To define the problem more precisely, we use the term ``normal
observers'' to refer to those that evolve as a result of
non-equilibrium processes that occur in the wake of the hot big
bang.  If our universe is approaching a stable de Sitter
spacetime, then the total number of normal observers that will
ever exist in a fixed comoving volume of the universe is finite. 
On the other hand, the cumulative number of Boltzmann brains
grows without bound over time, growing roughly as the volume,
proportional to $e^{3H_\Lambda t}$. When extracting the
predictions of this theory, such an infinite preponderance of
Boltzmann brains cannot be ignored. 

For example, suppose that some normal observer, at some moment in
her lifetime, tries to make a prediction about her next
observation.  According to the theory there would be an infinite
number of Boltzmann brains, distributed throughout the spacetime,
that would happen to share exactly all her memories and thought
processes at that moment. Since all her knowledge is shared with
this set of Boltzmann brains, for all she knows she could equally
likely be any member of the set.  The probability that she is a
normal observer is then arbitrarily small, and all predictions
would be based on the proposition that she is a Boltzmann brain. 
The theory would predict, therefore, that the next observations
that she will make, if she survives to make any at all, will be
totally incoherent, with no logical relationship to the world
that she thought she knew. (While it is of course true that some
Boltzmann brains might experience coherent observations, for
example by living in a Boltzmann solar system, it is easy to show
that Boltzmann brains with such dressing would be vastly
outnumbered by Boltzmann brains without any coherent
environment.) Thus, the continued orderliness of the world that
we observe is distinctly at odds with the predictions of a
Boltzmann-brain-dominated cosmology.\footnote{Here
we are taking a completely mechanistic view of the brain,
treating it essentially as a highly sophisticated computer. 
Thus, the normal observer and the Boltzmann brains can be thought
of as a set of logically equivalent computers running the same
program with the same data, and hence they behave identically
until they are affected by further input, which might be
different.  Since the computer program cannot determine whether
it is running inside the brain of one of the normal observers or
one of the Boltzmann brains, any intelligent probabilistic
prediction that the program makes about the next observation
would be based on the assumption that it is equally likely to be
running on any member of that set.}

This problem was recently addressed by Page~\cite{Page06}, who
concluded that the least unattractive way to produce more normal
observers than Boltzmann brains is to require that our vacuum
should be rather unstable.  More specifically, it should decay
within a few Hubble times of vacuum energy domination; that is,
in 20 billion years or so.

In the context of inflationary cosmology, however, this problem
acquires a new twist. Inflation is generically eternal, with the
physical volume of false-vacuum inflating regions increasing
exponentially with time and ``pocket universes'' like ours
constantly nucleating out of the false vacuum.  In an eternally
inflating multiverse, the numbers of normal observers and
Boltzmann brains produced over the course of eternal inflation
are both infinite.  They can be meaningfully compared only after
one adopts some prescription to regulate the infinities. 

The problem of regulating the infinities in an eternally
inflating multiverse is known as the measure
problem~\cite{MPreviews}, and has been under discussion for some
time.  It is crucially important in discussing predictions for
any kind of observation.  Most of the discussion, including the
discussion in this paper, has been confined to the classical
approximation.  While one might hope that someday there will be
an answer to this question based on a fundamental
principle~\cite{principle}, most of the work on this subject has
focussed on proposing plausible measures and exploring their
properties.  Indeed, a number of measures have been
proposed~\cite{Linde:1986fd,LM,LLM,GBLL,AV94,AV95,Linde06,Linde07,
GTV,pockets,GSPVW,ELM,diamond,censor,Vanchurin07,Winitzki08,LVW},
and some of them have already been disqualified, as they make
predictions that conflict with observations.

In particular, if one uses the proper-time cutoff measure
\cite{Linde:1986fd,LM,LLM,GBLL,AV94}, one
encounters the ``youngness paradox,'' predicting that humans
should have evolved at a very early cosmic time, when the
conditions for life were rather hostile~\cite{Tegmark:2004qd}. 
The youngness problem, as well as the Boltzmann brain problem,
can be avoided in the stationary measure
\cite{Linde07,LVW}, which is an improved version of the
proper-time cutoff measure. However, the stationary measure, as
well as the pocket-based measure, is afflicted with a runaway
problem, suggesting that we should observe extreme values (either
very small or very large) of the primordial density contrast
$Q$~\cite{FHW,QGV} and the gravitational constant $G$~\cite{GS},
while these parameters appear to sit comfortably in the middle of
their respective anthropic ranges~\citenosort{anthQ,GS}. Some
suggestions to get around this issue have been described in
Refs.~\cite{GarciaBellido:1994ci,QGV,Linde:2005yw,HWY}. In
addition, the pocket-based measure seems to suffer from the
Boltzmann brain problem. The comoving coordinate measure
\cite{Starobinsky:1986fx,Linde:1986fd} and the causal-patch
measures~\cite{diamond,censor} are free from these problems, but
have an unattractive feature of depending sensitively on the
initial state of the multiverse.  This does not seem to mix well
with the attractor nature of eternal inflation: the asymptotic
late-time evolution of an eternally inflating universe is
independent of the starting point, so it seems appealing for the
measure to maintain this property.  Since the scale-factor cutoff
measure\footnote{This measure is sometimes referred to as the
volume-weighted scale-factor cutoff measure, but we will define
it below in terms of the counting of events in spacetime, so the
concept of weighting will not be relevant.  The term
``volume-weighted'' is relevant when a measure is described as a
prescription for defining the probability distribution for the
value of a field.  In Ref.~\cite{Linde06}, this measure is called
the ``pseudo-comoving volume-weighted measure.''}
\cite{LM,LLM,GBLL,AV95,Linde06,Clifton} has been shown to be free
of all of the above issues~\cite{DSGSV}, we consider it to be a
promising candidate for the measure of the multiverse.

As we have indicated, the relative abundance of normal observers
and Boltzmann brains depends on the choice of measure over the
multiverse.  This means the predicted ratio of Boltzmann brains
to normal observers can be used as yet another criterion to
evaluate a prescription to regulate the diverging volume of the
multiverse: regulators that predict normal observers are greatly
outnumbered by Boltzmann brains should be ruled out.  This
criterion has been studied in the context of several multiverse
measures, including a causal patch measure~\cite{BF06}, several
measures associated with globally defined time
coordinates~\citenosort{Linde06,Page3,Linde07,BFY,LVW}, and the
pocket-based measure~\cite{AVBB}.  In this work, we apply this
criterion to the scale-factor cutoff measure, extending the
investigation that was initiated in Ref.~\cite{Linde06}.  We show
that the scale-factor cutoff measure gives a finite ratio of
Boltzmann brains to normal observers; if certain assumptions
about the landscape are valid, the ratio can be
small.\footnote{In a paper that appeared simultaneously with
version 1 of this paper, Raphael Bousso, Ben Freivogel, and
I-Sheng Yang independently analyzed the Boltzmann brain problem
for the scale-factor cutoff measure~\cite{BFY2}.}

The remainder of this paper is organized as follows.  In
Section~\ref{sec:sfcutoff}, we provide a brief description of the
scale-factor cutoff and describe salient features of the
multiverse under the lens of this measure.  In
Section~\ref{sec:BBabundance} we calculate the ratio of Boltzmann
brains to normal observers in terms of multiverse volume
fractions and transition rates.  The volume fractions are
discussed in Section~\ref{sec:minilandscapes}, in the context of
toy landscapes, and the section ends with a general description
of the conditions necessary to avoid Boltzmann brain domination.
The rate of Boltzmann brain production and the rate of vacuum
decay play central roles in our calculations, and these are
estimated in Section~\ref{sec:nucleationvsdecay}.  Concluding
remarks are provided in Section~\ref{sec:conclusions}.


\section{The Scale Factor Cutoff}
\label{sec:sfcutoff}

Perhaps the simplest way to regulate the infinities of eternal
inflation is to impose a cutoff on a hypersurface of constant
global time~\cite{LM,LLM,GBLL,AV94,AV95}.  One starts with a
patch of a spacelike hypersurface $\Sigma$ somewhere in an
inflating region of spacetime, and follows its evolution along
the congruence of geodesics orthogonal to $\Sigma$.  The
scale-factor time is defined as
\beq
t=\ln a \,,
\label{tdef}
\eeq
where $a$ is the expansion factor along the geodesics.  The
scale-factor time is related to the proper time $\tau$ by
\beq
dt = H\,d\tau \,,
\label{ttau}
\eeq
where $H$ is the Hubble expansion rate of the congruence.  The
spacetime region swept out by the congruence will typically
expand to unlimited size, generating an infinite number of
pockets.  (If the patch does not grow without limit, one chooses
another initial patch $\Sigma$ and starts again.) The resulting
four-volume is infinite, but we cut it off at some fixed
scale-factor time $t=t_c$.  To find the relative probabilities of
different events, one counts the numbers of such events in the
finite spacetime volume between $\Sigma$ and the $t=t_c$
hypersurface, and then takes the limit $t_c\to\infty$. 

The term ``scale factor'' is often used in the context of
homogeneous and isotropic geometries; yet on very large and on
very small scales the multiverse may be very inhomogeneous.  A
simple way to deal with this is to take the factor $H$ in
Eq.~(\ref{ttau}) to be the local divergence of the four-velocity
vector field along the congruence of geodesics orthogonal to
$\Sigma$,
\beq
H(x) \equiv (1/3) \,u^\mu_{\phantom{\mu};\,\mu} \,.
\label{localsft}
\eeq
When more than one geodesic passes through a point, the
scale-factor time at that point may be taken to be the smallest
value among the set of geodesics.  In collapsing regions $H(x)$
is negative, in which case the corresponding geodesics are
continued unless or until they hit a singularity.

This ``local'' definition of scale-factor time has a simple
geometric meaning.  The congruence of geodesics can be thought of
as representing a ``dust'' of test particles scattered uniformly
on the initial hypersurface $\Sigma$.  As one moves along the
geodesics, the density of the dust in the orthogonal plane
decreases.  The expansion factor $a$ in Eq.~(\ref{tdef}) can then
defined as $a\propto\rho^{-1/3}$, where $\rho$ is the density of
the dust, and the cutoff is triggered when $\rho$ drops below
some specified level.

Although the local scale-factor time closely follows the FRW
scale factor in expanding spacetimes --- such as inflating
regions and thermalized regions not long after reheating --- it
differs dramatically from the FRW scale factor as small-scale
inhomogeneities develop during matter domination in universes
like ours.  In particular, the local scale-factor time nearly
grinds to a halt in regions that have decoupled from the Hubble
flow.  It is not clear whether we should impose this particular
cutoff, which would essentially include the entire lifetime of
any nonlinear structure that forms before the cutoff, or impose a
cutoff on some nonlocal time variable that more closely tracks
the FRW scale factor.\footnote{The distinction between these two
forms of scale-factor time was first pointed out
by Bousso, Freivogel, and Yang in Ref.~\cite{BFY2}.}

There are a number of nonlocal modifications of scale factor time
that both approximate our intuitive notion of FRW averaging and
also extend into more complicated geometries.  One drawback of
the nonlocal approach is that no single choice looks more
plausible than the others. For instance, one nonlocal method is
to define the factor $H$ in Eq.~(\ref{ttau}) by spatial averaging
of the quantity $H(x)$ in Eq.~(\ref{localsft}).  A complete
implementation of this approach, however, involves many seemingly
arbitrary choices regarding how to define the hypersurfaces over
which $H(x)$ should be averaged, so we here set this possibility
aside.  A second, simpler method is to use the local scale-factor
time defined above, but to generate a new cutoff hypersurface by
excluding the future lightcones of all points on the original
cutoff hypersurface.  In regions with nonlinear inhomogeneities,
the underdense regions will be the first to reach the
scale-factor cutoff, after which they quickly trigger the cutoff
elsewhere.  The resulting cutoff hypersurface will not be a
surface of constant FRW scale factor, but the fluctuations of the
FRW scale factor on this surface should be insignificant.

As a third and final example of a nonlocal modification of scale
factor time, we recall the description of the local scale-factor
cutoff in terms the density $\rho$ of a dust of test particles. 
Instead of such a dust, consider a set of massless test
particles, emanating uniformly in all directions from each point
on the initial hypersurface $\Sigma$.  We can then construct the
conserved number density current $J^\mu$ for the gas of test
particles, and we can define $\rho$ as the rest frame number
density, i.e. the value of $J^0$ in the local Lorentz frame in
which $J^i=0$, or equivalently $\rho = \sqrt{J^2}$. Defining $a
\propto \rho^{-1/3}$, as we did for the dust of test particles,
we apply the cutoff when the number density $\rho$ drops below
some specified level.  Since null geodesics are barely perturbed
by structure formation, the strong perturbations inherent in the
local definition of scale factor time are avoided.  Nonetheless,
we have not studied the properties of this definition of scale
factor time, and they may lead to complications.  Large-scale
anisotropic flows in the gas of test particles can be generated
as the particles stream into expanding bubbles from outside.
Since the null geodesics do not interact with matter except
gravitationally, these anisotropies will not be damped in the
same way as they would be for photons.  The large-scale flow of
the gas will not redshift in the normal way, either; for example,
if the test particles in some region of an FRW universe have a
nonzero mean velocity relative to the comoving frame, the
expansion of the universe will merely reduce the energies of all
the test particles by the same factor, but will not cause the
mean velocity to decrease.  Thus, the detailed predictions for
this definition of scale-factor cutoff measure remain a matter
for future study.

The local scale-factor cutoff and each of the three nonlocal
definitions correspond to different global-time parameterizations
and thus to different spacetime measures.  In general they make
different predictions for physical observables; however with
regard to the relative number of normal observers and Boltzmann
brains, their predictions are essentially the same.  For the
remainder of this paper we refer to the generic nonlocal
definition of scale factor time, for which we take FRW time as a
suitable approximation.  Note that the use of local scale factor
time would make it slightly easier to avoid Boltzmann brain
domination, since it would increase the count of normal observers
while leaving the count of Boltzmann brains essentially
unchanged.

To facilitate later discussion, let us now describe some general
properties of the multiverse.  The volume fraction $f_i$ occupied
by vacuum $i$ on constant scale-factor time slices can be found
from the rate equation~\cite{recycling},
\beq
{df_i\over{dt}}= \sum_j M_{ij}f_j \,,
\label{rateeq}
\eeq
where the transition matrix $M_{ij}$ is given by
\beq
M_{ij}=\kappa_{ij}-\delta_{ij}\sum_r \kappa_{ri} \,,
\label{Mij}
\eeq
and $\kappa_{ij}$ is the transition rate from vacuum $j$ to
vacuum $i$ per Hubble volume per Hubble time.  This rate can also
be written
\beq
\kappa_{ij}= (4\pi/3)H_j^{-4} \Gamma_{ij} \,,
\label{kappa}
\eeq
where $\Gamma_{ij}$ is the bubble nucleation rate per unit
spacetime volume and $H_j$ is the Hubble expansion rate in
vacuum~$j$. 

The solution of Eq.~(\ref{rateeq}) can be written in terms of the
eigenvectors and eigenvalues of the transition matrix $M_{ij}$. 

It is easily verified that each terminal vacuum is an eigenvector
with eigenvalue zero.  We here define ``terminal vacua'' as those
vacua $j$ for which $\kappa_{ij}=0$ for all $i$. Thus the
terminal vacua include both negative-energy vacua, which collapse
in a big crunch, and stable zero-energy vacua. It was shown in
Ref.~\cite{GSPVW} that all of the other eigenvalues of $M_{ij}$
have negative real parts.  Moreover, the eigenvalue with the
smallest (by magnitude) real part is pure real; we call it the
``dominant eigenvalue'' and denote it by $-q$ (with $q>0$). 
Assuming that the landscape is irreducible, the dominant
eigenvalue is nondegenerate.  In that case the probabilities
defined by the scale-factor cutoff measure are independent of the
initial state of the multiverse, since they are determined by the
dominant eigenvector.\footnote{In this work we assume that the
multiverse is irreducible; that is, any metastable inflating
vacuum is accessible from any other such vacuum via a sequence of
tunneling transitions.  Our results, however, can still be
applied when this condition fails.  In that case the dominant
eigenvalue can be degenerate, in which case the asymptotic future
is dominated by a linear combination of dominant eigenvectors
that is determined by the initial state.  If transitions that
increase the vacuum energy density are included, then the
landscape can be reducible only if it splits into several
disconnected sectors.  That situation was discussed in Appendix A
of Ref.~\cite{DSGSV}, where two alternative prescriptions were
described.  The first prescription (preferred by the authors)
leads to initial-state dependence only if two or more sectors
have the same dominant eigenvalue $q$, while the second
prescription always leads to initial-state dependence.}

For an irreducible landscape, the late-time asymptotic solution
of Eq.~(\ref{rateeq}) can be written in the
form\footnote{$M_{ij}$ is not necessarily diagonalizable, but
Eq.~(\ref{asymptotic}) applies in any case.  It is always
possible to form a complete basis from eigenvectors and
generalized eigenvectors, where generalized eigenvectors satisfy
$(M-\lambda I)^k s = 0$, for $k > 1$.  The generalized
eigenvectors appear in the solution with a time dependence given
by $e^{\lambda t}$ times a polynomial in $t$.  These terms are
associated with the nonleading eigenvalues omitted from
Eq.~(\ref{asymptotic}), and the polynomials in $t$ will not
change the fact that they are nonleading.}
\beq
f_j(t)=f_j^{(0)} + s_j e^{-qt} + \ldots \,,
\label{asymptotic}
\eeq
where the constant term $f_j^{(0)}$ is nonzero only in terminal
vacua and $s_j$ is proportional to the eigenvector of $M_{ij}$
corresponding to the dominant eigenvalue $-q$, with the constant
of proportionality determined by the initial distribution of
vacua on $\Sigma$. It was shown in Ref.~\cite{GSPVW} that $s_j
\le 0$ for terminal vacua, and $s_j > 0$ for nonterminal vacua,
as is needed for Eq.~(\ref{asymptotic}) to describe a nonnegative
volume fraction, with a nondecreasing fraction assigned to any
terminal vacuum.

By inserting the asymptotic expansion (\ref{asymptotic}) into the
differential equation (\ref{rateeq}) and extracting the leading
asymptotic behavior for a nonterminal vacuum $i$, one can show
that
\beq
(\kappa_i - q) s_i = \sum_j \kappa_{ij} \, s_j \, ,
\label{kappa-q}
\eeq
where $\kappa_j$ is the total transition rate out of vacuum $j$,
\beq
\kappa_j \equiv \sum_i \kappa_{ij} \,.
\label{kappaj}
\eeq
The positivity of $s_i$ for nonterminal vacua then implies
rigorously that $q$ is less than the decay rate of the
slowest-decaying vacuum in the landscape:
\beq
q \le \kappa_{\rm min} \equiv {\rm min}\{\kappa_j\} \, .
\label{qbound}
\eeq

Since ``upward'' transitions (those that increase the energy
density) are generally suppressed, we can gain some intuition by
first considering the case in which all upward transition rates
are set to zero.  (Such a landscape is reducible, so the dominant
eigenvector can be degenerate.) In this case $M_{ij}$ is
triangular, and the eigenvalues are precisely the decay rates
$\kappa_i$ of the individual states.  The dominant eigenvalue $q$
is then exactly equal to $\kappa_{\rm min}$.

If upward transitions are included but assumed to have a very low
rate, then the dominant eigenvalue $q$ is approximately equal to
the decay rate of the slowest-decaying vacuum~\cite{Delia},
\beq
q\approx \kappa_{\rm min} \, .
\eeq
The slowest-decaying vacuum (assuming it is unique) is the one
that dominates the asymptotic late-time volume of the multiverse,
so we call it the dominant vacuum and denote it by $D$.  Hence,
\beq
q\approx \kappa_D \,.
\label{qD}
\eeq
The vacuum decay rate is typically exponentially suppressed, so
for the slowest-decaying vacuum we expect it to be extremely
small,
\beq
q \lll 1 \,.
\label{gammasmall}
\eeq
Note that the corrections to Eq.~(\ref{qD}) are comparable to the
upward transition rate from $D$ to higher-energy vacua, but for
large energy differences this transition rate is suppressed by
the factor $\exp(-8\pi^2/H_D^2)$~\cite{LeeWeinberg}.  Here and
throughout the remainder of this paper we use reduced Planck
units, where $8\pi G=c=k_B=1$. We shall argue in
Section~\ref{sec:nucleationvsdecay} that the dominant vacuum is
likely to have a very low energy density, so the correction to
Eq.~(\ref{qD}) is very small even compared to $q$.

A possible variant of this picture, with similar consequences,
could arise if one assumes that the landscape includes states
with nearby energy densities for which the upward transition rate
is not strongly suppressed.  In that case there could be a group
of vacuum states that undergo rapid transitions into each other,
but very slow transitions to states outside the group.  The role
of the dominant vacuum could then be played by this group of
states, and $q$ would be approximately equal to some
appropriately averaged rate for the decay of these states to
states outside the group.  Under these circumstances $q$ could be
much less than $\kappa_{\rm min}$.  An example of such a
situation is described in Subsection~~\ref{ssec:fn6}.

In the asymptotic limit of late scale-factor time $t$, the
physical volumes in the various nonterminal vacua are given by
\beq
V_j(t) = V_0\, s_j\, e^{(3-q)\,t} \,,
\label{Vj}
\eeq
where $V_0$ is the volume of the initial hypersurface $\Sigma$
and $e^{3t}$ is the volume expansion factor.  The volume growth
in Eq.~(\ref{Vj}) is (very slightly) slower than $e^{3t}$ due to
the constant loss of volume from transitions to terminal vacua. 
Note that even though upward transitions from the dominant vacuum
are strongly suppressed, they play a crucial role in populating
the landscape~\cite{Delia}.  Most of the volume in the asymptotic
solution of Eq.~(\ref{Vj}) originates in the dominant vacuum $D$,
and ``trickles'' to the other vacua through a series of
transitions starting with at least one upward jump. 


\section{The Abundance of Normal Observers and Boltzmann Brains}
\label{sec:BBabundance}

Let us now calculate the relative abundances of Boltzmann brains
and normal observers, in terms of the vacuum transition rates and
the asymptotic volume fractions. 

Estimates for the numerical values of the Boltzmann brain
nucleation rates and vacuum decay rates will be discussed in
Section \ref{sec:nucleationvsdecay}, but it is important at this
stage to be aware of the kind of numbers that will be considered. 
We will be able to give only rough estimates of these rates, but
the numbers that will be mentioned in Section
\ref{sec:nucleationvsdecay} will range 
from $\exp\,(-10^{120})$ to $\exp\,(-10^{16})$. Thus, when we
calculate the ratio $\N^{\rm BB}/\N^{\rm NO}$ of Boltzmann brains
to normal observers, the natural logarithm of this ratio will
always include one term with a magnitude of at least $10^{16}$. 
Consequently, the presence or absence of any term in $\ln
(\N^{\rm BB}/\N^{\rm NO})$ that is small compared to $10^{16}$ is
of no relevance.  We therefore refer to any factor $f$ for which
\beq
|\ln f | < 10^{14}
  \label{orderone}
\eeq
as ``roughly of order one.'' In the calculation of $\N^{\rm
BB}/\N^{\rm NO}$ such factors --- although they may be minuscule
or colossal by ordinary standards --- can be ignored. It will not
be necessary to keep track of factors of 2, $\pi$, or even
$10^{10^8}$.  Dimensionless coefficients, factors of $H$, and
factors coming from detailed aspects of the geometry are
unimportant, and in the end all of these will be ignored. We
nonetheless include some of these factors in the intermediate
steps below simply to provide a clearer description of the
calculation.

We begin by estimating the number of normal observers that will
be counted in the sample spacetime region specified by the
scale-factor cutoff measure. Normal observers arise during the
big bang evolution in the aftermath of slow-roll inflation and
reheating.  The details of this evolution depend not only on the
vacuum of the pocket in question, but also on the parent vacuum
from which it nucleated~\cite{Aguirre}.  That is, if we view each
vacuum as a local minimum in a multidimensional field space, then
the dynamics of inflation in general depend on the direction from
which the field tunneled into the local minimum.  We therefore
label pockets with two indices, $ik$, indicating the pocket and
parent vacua respectively.

To begin, we restrict our attention to a single ``anthropic''
pocket --- i.e., one that produces normal observers --- which
nucleates at scale-factor time $t_{\rm nuc}$.  The internal
geometry of the pocket is that of an open FRW universe.  We
assume that, after a brief curvature-dominated period
$\Delta\tau\sim H_k^{-1}$, slow-roll inflation inside the pocket
gives $N_e$ e-folds of expansion before thermalization. 
Furthermore, we assume that all normal observers arise at a fixed
number $N_O$ of e-folds of expansion after thermalization.  (Note
that $N_e$ and $N_O$ are both measured along FRW comoving
geodesics inside the pocket, which do not initially coincide
with, but rapidly asymptote to, the ``global'' geodesic
congruence that originated outside the pocket.) We denote the
fixed-internal-time hypersurface on which normal observers arise
by $\Sigma^{\rm NO}$, and call the average density of observers
on this hypersurface $n^{\rm NO}_{ik}$.

The hypersurface $\Sigma^{\rm NO}$ would have infinite volume,
due to the constant expansion of the pocket, but this divergence
is regulated by the scale-factor cutoff $t_c$, because the global
scale-factor time $t$ is not constant over the $\Sigma^{\rm NO}$
hypersurface.  For the pocket described above, the regulated
physical volume of $\Sigma^{\rm NO}$ can be written as
\beq
V^{(ik)}_O(t_{\rm nuc})=H_k^{-3}e^{3(N_e+N_O)}\, w(t_c-t_{\rm
nuc}-N_e-N_O) \,,
\label{Vik}
\eeq
where the exponential gives the volume expansion factor coming
from slow-roll inflation and big bang evolution to the
hypersurface $\Sigma^{\rm NO}$, and $H_k^{-3} w(t_c-t_{\rm
nuc}-N_e-N_O)$ describes the comoving volume of the part of the
$\Sigma^{\rm NO}$ hypersurface that is underneath the cutoff. The
function $w(t)$ was calculated, for example, in Refs.~\cite{VW97}
and \cite{BFY}, and is applied to scale-factor cutoff measure in
Ref.~\cite{DSGSV2}.  Its detailed form will not be needed to
determine the answer up to a factor that is roughly of order one,
but to avoid mystery we mention that $w(t)$ can be written as
\beq
w(t) = {\pi \over 2} \int_0^{\bar\xi(t)} \, \sinh^2(\xi) \, d \xi
= {\pi \over 8} \left[ \sinh\bigl(2\bar\xi(t)\bigr) - 2
\bar\xi(t) \right] \, ,
\label{woft}
\eeq
where $\bar\xi(t_c-t_{\rm nuc}-N_e-N_O)$ is the maximum value of
the Robertson-Walker radial coordinate $\xi$ that lies under the
cutoff. If the pocket universe begins with a moderate period of
inflation ($\exp(N_e) \gg 1$) with the same vacuum energy as
outside, then
\beq
\bar\xi(t) \approx 2 \cosh^{-1} \left( e^{t / 2} \right) \,.
\label{xibar}
\eeq

Eq.~(\ref{Vik}) gives the physical volume on the $\Sigma^{\rm
NO}$ hypersurface for a single pocket of type $ik$, which
nucleates at time $t_{\rm nuc}$. The number of $ik$-pockets that
nucleate between time $t_{\rm nuc}$ and $t_{\rm nuc}+dt_{\rm
nuc}$ is
\begin{eqnarray}
dn_{\rm nuc}^{(ik)}(t_{\rm nuc}) &=& (3/4\pi)H_k^3\kappa_{ik}\,
V_k(t_{\rm nuc})\, dt_{\rm nuc} \nonumber \\
&=& (3/4\pi)H_k^3\kappa_{ik}s_k\,V_0\,e^{(3-q)\,{t_{\rm
nuc}}}\,dt_{\rm nuc} \,,
\end{eqnarray}
where we use Eq.~(\ref{Vj}) to give $V_k(t_{\rm nuc})$. The total
number of normal observers in the sample region is then
\begin{eqnarray}
\N_{ik}^{\rm NO} &=& n^{\rm NO}_{ik}\, \int^{t_c-N_e-N_O}\! 
V_O^{(ik)}(t_{\rm nuc})\,dn_{\rm nuc}^{(ik)}(t_{\rm nuc})\nonumber\\
& \approx  & n_{ik}^{\rm NO}\kappa_{ik}s_k\,V_0\,e^{(3-q)\,t_c}\! 
\int_0^\infty\! w(z)\,e^{-(3-q)\,z}\, dz \,.\quad
\end{eqnarray}
In the first expression we have ignored the (very small)
probability that pockets of type $ik$ may transition to other
vacua during slow-roll inflation or during the subsequent period
$N_O$ of big bang evolution.  In the second line, we have changed
the integration variable to $z=t_c-t_{\rm nuc}-N_e-N_O$
(reversing the direction of integration) and have dropped the
${\mathcal O}(1)$ prefactors, and also the factor
$e^{q(N_e+N_O)}$, since $q$ is expected to be extraordinarily
small.  We have kept $e^{-q t_c}$, since we are interested in the
limit $t_c \to
\infty$.  We have also kept the factor $e^{q z}$ long enough to
verify that the integral converges with or without the factor, so
we can carry out the integral using the approximation $q \approx
0$, resulting in an ${\mathcal O}(1)$ prefactor that we will
drop.

Finally,
\beq
\N_{ik}^{\rm NO} \approx n_{ik}^{\rm NO}\kappa_{ik}\,
s_k\,V_0\,e^{(3-q)\,t_c} \ .
\label{Nreg}
\eeq
Note that the expansion factor $e^{3(N_e+N_O)}$ in
Eq.~(\ref{Vik}) was canceled when we integrated over nucleation
times, illustrating the mild youngness bias of the scale-factor
cutoff measure.  The expansion of the universe is canceled, so
objects that form at a certain density per physical volume in the
early universe will have the same weight as objects that form at
the same density per physical volume at a later time, despite the
naive expectation that there is more volume at later times.

To compare, we now need to calculate the number of Boltzmann
brains that will be counted in the sample spacetime region.
Boltzmann brains can be produced in any anthropic vacuum, and
presumably in many non-anthropic vacua as well. Suppose Boltzmann
brains are produced in vacuum $j$ at a rate $\Gamma_j^{\rm BB}$
per unit spacetime volume.  The number of Boltzmann brains
$\N_j^{\rm BB}$ is then proportional to the total four-volume in
that vacuum.  Imposing the cutoff at scale-factor time $t_c$,
this four-volume is
\begin{eqnarray}
{\mathcal V}_j^{(4)} &=&\int^{t_c} V_j(t)\,d\tau \,=\, H_j^{-1}
\int^{t_c}V_j(t)\,dt \nonumber \\ 
&=& {1\over{3-q}}H_j^{-1}s_j \, V_0 \, e^{(3-q)\,t_c} \,,
\end{eqnarray}
where we have used Eq.~(\ref{Vj}) for the asymptotic volume
fraction. By setting $d\tau = H_j^{-1} \, d t$, we have ignored
the time-dependence of $H(\tau)$ in the earlier stages of cosmic
evolution, assuming that only the late-time de Sitter evolution
is relevant.  In a similar spirit, we will assume that the
Boltzmann brain nucleation rate $\Gamma_j^{\rm BB}$ can be
treated as time-independent, so the total number of Boltzmann
brains nucleated in vacua of type $j$, within the sample volume,
is given by
\beq
\N_j^{\rm BB} \,\approx\, \Gamma_j^{\rm BB} H_j^{-1}
s_j \, V_0 \, e^{(3-q)\,t_c} \,,
\label{Nfreak}
\eeq
where we have dropped the ${\mathcal O}(1)$ numerical factor.

For completeness, we may want to consider the effects of early
universe evolution on Boltzmann brain production, effects which
were ignored in Eq.~(\ref{Nfreak}).  We will separate the effects
into two categories: the effects of slow-roll inflation at the
beginning of a pocket universe, and the effects of reheating.

To account for the effects of slow-roll inflation, we argue that,
within the approximations used here, there is no need for an
extra calculation.  Consider, for example, a pocket universe $A$
which begins with a period of slow-roll inflation during which
$H(\tau) \approx H_{\rm slow\ roll} = \hbox{const}.$ Consider
also a pocket universe $B$, which throughout its evolution has $H
= H_{\rm slow\ roll}$, and which by hypothesis has the same
formation rate, Boltzmann brain nucleation rate, and decay rates
as pocket $A$.  Then clearly the number of Boltzmann brains
formed in the slow roll phase of pocket $A$ will be smaller than
the number formed throughout the lifetime of pocket $B$.  Since
we will require that generic bubbles of type $B$ do not
overproduce Boltzmann brains, there will be no need to worry
about the slow-roll phase of bubbles of type $A$.

To estimate how many Boltzmann brains might form as a consequence
of reheating, we can make use of the calculation for the
production of normal observers described above. We can assume
that the Boltzmann brain nucleation rate has a spike in the
vicinity of some particular hypersurface in the early universe,
peaking at some value $\Gamma^{\rm BB}_{{\rm reheat},ik}$ which
persists roughly for some time interval $\Delta \tau^{\rm
BB}_{{\rm reheat},ik}$, producing a density of Boltzmann brains
equal to $\Gamma^{\rm BB}_{{\rm reheat},ik} \, \Delta \tau^{\rm
BB}_{{\rm reheat},ik}$.  This spatial density is converted into a
total number for the sample volume in exactly the same way that
we did for normal observers, leading to
\beq
\N_{ik}^{\rm BB,reheat} \approx \Gamma^{\rm BB}_{{\rm
reheat},ik} \, \Delta \tau^{\rm BB}_{{\rm reheat},ik} \,
\kappa_{ik}\, s_k\,V_0\,e^{(3-q)\,t_c} \ .
\label{Nreheat}
\eeq
Thus, the dominance of normal observers is assured if
\beq
\sum_{i,k} \, \Gamma^{\rm BB}_{{\rm reheat},ik} \, \Delta
\tau^{\rm BB}_{{\rm reheat},ik} \kappa_{ik}\, s_k \ll \sum_{i,k}
\, n_{ik}^{\rm NO} \kappa_{ik}\, s_k \, . 
\label{Domreheat}
\eeq
If Eq.~(\ref{Domreheat}) did not hold, it seems likely that we
would suffer from Boltzmann brain problems regardless of our
measure.  We leave numerical estimates for Section
\ref{sec:nucleationvsdecay}, but we will see that Boltzmann brain
production during reheating is not a danger.

Ignoring the Boltzmann brains that form during reheating, the
ratio of Boltzmann brains to normal observers can be found by
combining Eqs.~(\ref{Nreg}) and (\ref{Nfreak}), giving
\beq
{\N^{\rm BB}\over{\N^{\rm NO}}} \approx
\frac{\sum_jH_j^3\kappa^{\rm BB}_j\,s_j}
{\sum_{i,\,k}n_{ik}^{\rm NO}\kappa_{ik}\,s_k} \,,
\label{BBregratio1}
\eeq
where the summation in the numerator covers only the vacua in
which Boltzmann brains can arise, the summation over $i$ in the
denominator covers only anthropic vacua, and the summation over
$k$ includes all of their possible parent vacua.  $\kappa^{\rm
BB}_j$ is the dimensionless Boltzmann brain nucleation rate in
vacuum $j$, related to $\Gamma^{\rm BB}_j$ by Eq.~(\ref{kappa}).
The expression can be further simplified by dropping the factors
of $H_j$ and $n_i^{\rm NO}$, which are roughly of order one, as
defined by Eq.~(\ref{orderone}).  We can also replace the sum
over $j$ in the numerator by the maximum over $j$, since the sum
is at least as large as the maximum term and no larger than the
maximum term times the number of vacua.  Since the number of
vacua (perhaps $10^{500}$) is roughly of order one, the sum over
$j$ is equal to the maximum up to a factor that is roughly of
order one.  We similarly replace the sum over $i$ in the
denominator by its maximum, but we choose to leave the sum over
$k$. Thus we can write
\beq
{\N^{\rm BB}\over{\N^{\rm NO}}} \sim
\frac{\max_j \{\kappa^{\rm BB}_j\,s_j\}}{\max_i 
\left\{\sum_k \kappa_{ik}\,s_k \right\}} \,,
\label{BBregratio2}
\eeq
where the sets of $j$ and $i$ are restricted as for
Eq.~(\ref{BBregratio1}). 

In dropping $n_i^{\rm NO}$, we are assuming that $n_i^{\rm NO}
H_i^3$ is roughly of order one, as defined at the beginning of
this section.  It is hard to know what a realistic value for
$n_i^{\rm NO} H_i^3$ might be, as the evolution of normal
observers may require some highly improbable events.  For
example, it was argued in Ref.~\cite{Koonin} that the probability
for life to evolve in a region of the size of our observable
universe per Hubble time may be as low as $\sim 10^{-1000}$.  But
even the most pessimistic estimates cannot compete with the small
numbers appearing in estimates of the Boltzmann brain nucleation
rate, and hence by our definition they are roughly of order one. 
Nonetheless, it is possible to imagine vacua for which $n_i^{\rm
NO}$ might be negligibly small, but still nonzero.  We shall
ignore the normal observers in these vacua; for the remainder of
this paper we will use the phrase ``anthropic vacuum'' to refer
only to those vacua for which $n_i^{\rm NO} H_i^3$ is roughly of
order one.

For any landscape that satisfies Eq.~(\ref{asymptotic}), which
includes any irreducible landscape, Eq.~(\ref{BBregratio2}) can
be simplified by using Eq.~(\ref{kappa-q}):
\beq
{\N^{\rm BB}\over{\N^{\rm NO}}} \sim
\frac{\max_j \{\kappa^{\rm BB}_j\,s_j\}}
{\max_i \{(\kappa_{i}-q)\,s_i\}} \,,
\label{BBregratio3}
\eeq
where the numerator is maximized over all vacua $j$ that support
Boltzmann brains, and the denominator is maximized over all
anthropic vacua $i$.

In order to learn more about the ratio of Boltzmann brains to
normal observers, we need to learn more about the volume
fractions $s_j$, a topic that will be pursued in the next
section.


\section{Mini-landscapes and the General Conditions to Avoid 
Boltzmann Brain Domination}
\label{sec:minilandscapes}

In this section we study a number of simple models of the
landscape, in order to build intuition for the volume fractions
that appear in Eqs.~(\ref{BBregratio2}) and~(\ref{BBregratio3}). 
The reader uninterested in the details may skip the pedagogical
examples given in Subsections \ref{ssec:FIB}--\ref{ssec:fn6}, and
continue with Subsection~\ref{ssec:genconds}, where we state the
general conditions that must be enforced in order to avoid
Boltzmann brain domination.


\subsection{The FIB Landscape}
\label{ssec:FIB}

Let us first consider a very simple model of the landscape,
described by the schematic
\beq
F\rightarrow I\rightarrow B \,,
\label{FIB}
\eeq
where $F$ is a high-energy false vacuum, $I$ is a positive-energy
anthropic vacuum, and $B$ is a terminal vacuum.  This model,
which we call the FIB landscape, was analyzed
in~Ref.~\cite{GSPVW} and was discussed in relation to the
abundance of Boltzmann brains in Ref.~\cite{Linde06}. As in
Ref.~\cite{Linde06}, we assume that both Boltzmann brains and
normal observers reside only in vacuum $I$. 

Note that the FIB landscape ignores upward transitions from $I$
to $F$.  The model is constructed in this way as an initial first
step, and also in order to more clearly relate our analysis to
that of Ref.~\cite{Linde06}.  Although the rate of upward
transitions is exponentially suppressed relative the other rates,
its inclusion is important for the irreducibility of the
landscape, and hence the nondegeneracy of the dominant eigenvalue
and the independence of the late-time asymptotic behavior from
the initial conditions of the multiverse. The results of this
subsection will therefore not always conform to the expectations
outlined in Section~\ref{sec:sfcutoff}, but this shortcoming is
corrected in the next subsection and all subsequent work in this
paper.

We are interested in the eigenvectors and eigenvalues of the rate
equation, Eq.~(\ref{rateeq}). In the FIB landscape the rate
equation gives
\begin{eqnarray}
\begin{array}{rl}
{\dot f}_F \! &= \, -\kappa_{IF} f_F  \\
\noalign{\vskip 4pt}
{\dot f}_I \! &= \, -\kappa_{BI}f_I +\kappa_{IF}f_F \,.
\end{array}
\label{rateFIB}
\end{eqnarray}
We ignore the volume fraction in the terminal vacuum as it is not
relevant to our analysis.  Starting with the ansatz,
\beq
{\bf f}(t) = {\bf s}\, e^{-qt} \,,
\eeq
we find two eigenvalues of Eqs.~(\ref{rateFIB}).  These are, with
their corresponding eigenvectors,
\begin{eqnarray}
\begin{array}{ll}
q_1=\kappa_{IF}\,, \quad & {\bf s}_1=(1,C)\,, \\ 
\noalign{\vskip 3pt}
q_2=\kappa_{BI}\,, \quad & {\bf s}_2 =(0,1)\,,
\end{array}
\label{solnFIB}
\end{eqnarray}
where the eigenvectors are written in the basis ${\bf
s}\equiv(s_F,s_I)$ and
\beq
C= {\kappa_{IF}\over{\kappa_{BI}-\kappa_{IF}}} \,. 
\eeq

Suppose that we start in the false vacuum $F$ at $t=0$, i.e. 
${\bf f}(t=0)=(1,0)$.  Then the solution of the FIB rate
equation, Eq.~(\ref{rateFIB}), is
\begin{eqnarray}
\begin{array}{rl}
f_F(t)\! &= \, e^{-\kappa_{IF}t}\,, \\
\noalign{\vskip4pt}
f_I(t)\! &= \, C \left( e^{-\kappa_{IF}t} -e^{-\kappa_{BI}t}\right) \,.
\end{array}
\label{fjFIB}
\end{eqnarray}
The asymptotic evolution depends on whether $\kappa_{IF} <
\kappa_{BI}$ (case I) or not (case II).  In case I,
\beq
{\bf f}(t\to\infty)={\bf s}_1 e^{- \kappa_{IF} t} \qquad
(\kappa_{IF} < \kappa_{BI}) \,,
\label{caseI}
\eeq
where ${\bf s}_1$ is given in Eq.~(\ref{solnFIB}), while in case
II
\beq
{\bf f}(t\to\infty)=\left( e^{- \kappa_{IF} t}~,|C|e^{-
\kappa_{BI} t}\right) \qquad (\kappa_{BI} < \kappa_{IF}) \,.
\label{caseII}
\eeq
In the latter case, the inequality of the rates of decay for the
two volume fractions arises from the reducibility of the FIB
landscape, stemming from our ignoring upward transitions from $I$
to $F$. 

For case I ($\kappa_{IF} < \kappa_{BI}$), we find the ratio of
Boltzmann brains to normal observers by evaluating
Eq.~(\ref{BBregratio2}) for the asymptotic behavior described by
Eq.~(\ref{caseI}):
\beq
{\N^{\rm BB} \over{\N^{\rm NO}}} \sim
\frac{\kappa^{\rm BB } s_I}{\kappa_{IF} s_F} \sim
\frac{\kappa^{\rm BB }}{\kappa_{IF}}
\frac{\kappa_{IF}}{\kappa_{BI}-\kappa_{IF}}
\sim {\kappa^{\rm BB} \over{\kappa_{BI}}} \,,
\eeq
where we drop $\kappa_{IF}$ compared to $\kappa_{BI}$ in the
denominator, as we are only interested in the overall scale of
the solution.  We find that the ratio of Boltzmann brains to
normal observers is finite, depending on the relative rate of
Boltzmann brain production to the rate of decay of vacuum $I$. 
Meanwhile, in case II (where $\kappa_{BI} < \kappa_{IF}$) we find
\beq
{\N^{\rm BB} \over{\N^{\rm NO}}} \sim
\frac{\kappa^{\rm BB}}{\kappa_{IF}} \,
e^{(\kappa_{IF}-\kappa_{BI})t} \to\infty\,.
\eeq
In this situation, the number of Boltzmann brains overwhelms the
number of normal observers; in fact the ratio diverges with time.

The unfavorable result of case II stems from the fact that, in
this case, the volume of vacuum $I$ grows faster than that of
vacuum $F$.  Most of this $I$-volume is in large pockets that
formed very early; and this volume dominates because the
$F$-vacuum decays faster than $I$, and is not replenished due to
the absence of upward transitions. This leads to Boltzmann brain
domination, in agreement with the conclusion reached in
Ref.~\cite{Linde06}.  Thus, the FIB landscape analysis suggests
that Boltzmann brain domination can be avoided only if the decay
rate of the anthropic vacuum is larger than both the decay rate
of its parent false vacuum $F$ and the rate of Boltzmann brain
production.  Moreover, the FIB analysis suggests that Boltzmann
brain domination in the multiverse can be avoided only if the
first of these conditions is satisfied for all vacua in which
Boltzmann brains exist.  This is a very stringent requirement,
since low-energy vacua like $I$ typically have lower decay rates
than high-energy vacua (see Section~\ref{sec:nucleationvsdecay}). 
We shall see, however, that the above conditions are
substantially relaxed in more realistic landscape models.


\subsection{The FIB Landscape with Recycling}

The FIB landscape of the preceding section is reducible, since
vacuum $F$ cannot be reached from vacuum $I$.  We can make it
irreducible by simply allowing upward transitions,
\beq
F\leftrightarrow I\rightarrow B \,.
\label{recFIB}
\eeq
This ``recycling FIB'' landscape is more realistic than the
original FIB landscape, because upward transitions out of
positive-energy vacua are allowed in semi-classical quantum
gravity~\cite{LeeWeinberg}.  The rate equation of the recycling
FIB landscape gives the eigenvalue system,
\begin{eqnarray}
\begin{array}{rl}
-qs_F\! &= \, -\kappa_{IF} s_F +\kappa_{FI}s_I\,, \\
\noalign{\vskip 4pt}
-qs_I\! &= \, -\kappa_{I}s_I  +\kappa_{IF}s_F\,,
\label{raterecFIB}
\end{array}
\end{eqnarray}
where $\kappa_I\equiv \kappa_{BI}+\kappa_{FI}$ is the total decay
rate of vacuum $I$, as defined in Eq.~(\ref{kappaj}). Thus, the
eigenvalues $q_1$ and $q_2$ correspond to the roots of
\beq
(\kappa_{IF}-q)(\kappa_{I}-q)=\kappa_{IF}\kappa_{FI}\,. 
\eeq

Further analysis is simplified if we note that upward transitions
from low-energy vacua like ours are very strongly suppressed,
even when compared to the other exponentially suppressed
transition rates, i.e.  $\kappa_{FI} \ll \kappa_{IF},
\kappa_{BI}$.  We are interested mostly in how this small
correction modifies the dominant eigenvector in the case where
$\kappa_{BI} < \kappa_{IF}$ (case II), which led to an infinite
ratio of Boltzmann brains to normal observers.  To the lowest
order in $\kappa_{FI}$, we find
\beq
q\approx \kappa_{I}-{\kappa_{IF}\kappa_{FI}
\over{\kappa_{IF}-\kappa_{I}}} \,,
\eeq
and
\beq
s_I\approx {{\kappa_{IF}-\kappa_{I}}\over{\kappa_{FI}}}s_F \gg
s_F\,.
\label{siFIB}
\eeq

The above equation is a consequence of the second of
Eqs.~(\ref{raterecFIB}), but it also follows directly from
Eq.~(\ref{kappa-q}), which holds in any irreducible landscape. In
this case $f_I(t)$ and $f_F(t)$ have the same asymptotic time
dependence, $\propto e^{-qt}$, so the ratio $f_I(t)/f_F(t)$
approaches a constant limit, $s_I/s_F \equiv R$. However, due to
the smallness of $\kappa_{FI}$, this ratio is extremely large. 
Note that the ratio of Boltzmann brains to normal observers is
proportional to $R$.  Although it is also proportional to the
minuscule Boltzmann brain nucleation rate (estimated in
Section~\ref{sec:nucleationvsdecay}), the typically huge value of
$R$ will still lead to Boltzmann brain domination (again, see
Section~\ref{sec:nucleationvsdecay} for relevant details).  But
the story is not over, since the recycling FIB landscape is still
far from realistic.


\subsection{A More Realistic Landscape}
\label{ssec:morerealistic}

In the recycling model of the preceding section, the anthropic
vacuum $I$ was also the dominant vacuum, while in a realistic
landscape this is not likely to be the case.  To see how it
changes the situation to have a non-anthropic vacuum as the
dominant one, we consider the model
\beq
A\leftarrow D\leftrightarrow F\rightarrow I\rightarrow B \,,
\eeq
which we call the ``ADFIB landscape.'' Here, $D$ is the dominant
vacuum and $A$ and $B$ are both terminal vacua.  The vacuum $I$
is still an anthropic vacuum, and the vacuum $F$ has large,
positive vacuum energy.  As explained in
Section~\ref{sec:nucleationvsdecay}, the dominant vacuum is
likely to have very small vacuum energy; hence we consider that
at least one upward transition (here represented as the
transition to $F$) is required to reach an anthropic vacuum. 

Note that the ADFIB landscape ignores the upward transition rate
from vacuum $I$ to $F$; however this is exponentially suppressed
relative the other transition rates pertinent to $I$ and, unlike
the situation in Subsection~\ref{ssec:FIB}, ignoring the upward
transition does not significantly affect our results. The
important property is that all vacuum fractions have the same
late-time asymptotic behavior, and this property is assured
whenever there is a unique dominant vacuum, and all inflating
vacua are accessible from the dominant vacuum via a sequence of
tunneling transitions.  The uniformity of asymptotic behaviors is
sufficient to imply Eq.~(\ref{kappa-q}), which implies
immediately that
\beq
\frac{s_I}{s_F}={\kappa_{IF} \over {\kappa_{BI} - q}} \approx
{\kappa_{IF}\over{\kappa_{BI}-\kappa_D}} \approx
{\kappa_{IF}\over\kappa_{BI}} \, ,
\label{sIsF}
\eeq
where we used $q \approx \kappa_D\equiv
\kappa_{AD}+\kappa_{FD}$, and assumed that $\kappa_D \ll \kappa_{BI}$.

This holds even if the decay rate of the anthropic vacuum $I$ is
smaller than that of the false vacuum $F$.

Even though the false vacuum $F$ may decay rather quickly, it is
constantly being replenished by upward transitions from the
slowly-decaying vacuum $D$, which overwhelmingly dominates the
physical volume of the multiverse.  Note that, in light of these
results, our constraints on the landscape to avoid Boltzmann
brain domination are considerably relaxed.  Specifically, it is
no longer required that the anthropic vacua decay at a faster
rate than their parent vacua.  Using Eq.~(\ref{sIsF}) with
Eq.~(\ref{BBregratio2}), the ratio of Boltzmann brains to normal
observers in vacuum $I$ is found to be
\beq
{\N_I^{\rm BB} \over{\N_I^{\rm NO}}} \sim
\frac{\kappa_I^{\rm BB } s_I}{\kappa_{IF} s_F} 
\sim {\kappa_I^{\rm BB} \over{\kappa_{BI}}} \,.
\label{BBratioC}
\eeq

If Boltzmann brains can also exist in the dominant vacuum $D$,
then they are a much more severe problem.  By applying
Eq.~(\ref{kappa-q}) to the $F$ vacuum, we find
\beq
\frac{s_F}{s_D} = {\kappa_{FD} \over \kappa_F - q} \approx
{\kappa_{FD} \over \kappa_F - \kappa_D} \approx {\kappa_{FD}
\over \kappa_F} \, ,
\label{sFsD}
\eeq
where $\kappa_F=\kappa_{IF}+\kappa_{DF}$, and where we have
assumed that $\kappa_D \ll \kappa_F$. The ratio of Boltzmann
brains in vacuum $D$ to normal observers in vacuum $I$ is then
\beq
{\N_D^{\rm BB} \over{\N_I^{\rm NO}}} \sim
\frac{\kappa_D^{\rm BB } s_D}{\kappa_{IF} s_F} 
\sim {\kappa_D^{\rm BB} \over{\kappa_{FD}}} {\kappa_F \over
\kappa_{IF}} \,. 
\eeq
Since we expect that the dominant vacuum has very small vacuum
energy, and hence a heavily suppressed upward transition rate
$\kappa_{FD}$, the requirement that ${\N_D^{\rm BB} / {\N_I^{\rm
NO}}}$ be small could be a very stringent one. Note that compared
to $s_D$, both $s_F$ and $s_I$ are suppressed by the small factor
$\kappa_{FD}$; however the ratio $s_I/s_F$ is independent of this
factor. 

Since $s_D$ is so large, one should ask whether Boltzmann brain
domination can be more easily avoided by allowing vacuum $D$ to
be anthropic.  The answer is no, because the production of normal
observers in vacuum $D$ is proportional (see Eq.~(\ref{Nreg})) to
the rate at which bubbles of $D$ nucleate, which is not large. 
$D$ dominates the spacetime volume due to slow decay, not rapid
nucleation.  If we assume that $D$ is anthropic and restrict
Eq.~(\ref{BBregratio2}) to vacuum $D$, we find using
Eq.~(\ref{sFsD}) that
\beq
{\N_D^{\rm BB} \over{\N_D^{\rm NO}}} \sim \frac{\kappa_D^{\rm BB
} s_D}{\kappa_{DF} s_F} \sim {\kappa_D^{\rm BB}
\over{\kappa_{FD}}} {\kappa_F \over \kappa_{DF}} \, ,
\label{DtoD}
\eeq
so again the ratio is enhanced by the extremely small upward
tunneling rate $\kappa_{FD}$ in the denominator.

Thus, in order to avoid Boltzmann brain domination, it seems we
have to impose two requirements: (1) the Boltzmann brain
nucleation rate in the anthropic vacuum $I$ must be less than the
decay rate of that vacuum, and (2) the dominant vacuum $D$ must
either not support Boltzmann brains at all, or must produce them
with a dimensionless rate $\kappa^{\rm BB}_D$ that is small even
compared to the upward tunneling rate $\kappa_{FD}$. If the
vacuum $D$ is anthropic then it must support Boltzmann brains,
so the domination by Boltzmann brains could be avoided only by
the stringent requirement $\kappa^{\rm BB}_{D} \ll \kappa_{FD}$.


\subsection{A Further Generalization}
\label{ssec:furthergen}

The conclusions of the last subsection are robust to more general
considerations.  To illustrate, let us generalize the ADFIB
landscape to one with many low-vacuum-energy pockets, described
by the schematic
\beq
A\leftarrow D\leftrightarrow F_j\rightarrow I_i\rightarrow B \,,
\label{biglandscape}
\eeq
where each high energy false vacuum $F_j$ decays into a set of
vacua $\{I_i\}$, all of which decay (for simplicity) to the same
terminal vacuum $B$.  The vacua $I_i$ are taken to be a large set
including both anthropic vacua and vacua that host only Boltzmann
brains.  Eq.~(\ref{kappa-q}) continues to apply, so
Eqs.~(\ref{sIsF}) and (\ref{sFsD}) are easily generalized to this
case, giving
\beq
s_{I_i} \approx {1 \over \kappa_{I_i}} \sum_j \kappa_{I_i F_j}
     \, s_{F_j}
\eeq
and
\beq
s_{F_j} \approx {1 \over \kappa_{F_j}} \kappa_{F_j D} \, s_D \, ,
\label{sFjratio}
\eeq
where we have assumed that $q \ll \kappa_{I_i}, \kappa_{F_j},$ as
we expect for vacua other than the dominant one. Using these
results with Eq.~(\ref{BBregratio2}), the ratio of Boltzmann
brains in vacua $I_i$ to normal observers in vacua $I_i$ is given
by
\begin{eqnarray}
{\N^{\rm BB}_{\{I_i\}}\over{\N^{\rm NO}_{\{I_i\}}}} &\sim&
\frac{ \max_i \left\{{\displaystyle \kappa^{\rm BB}_{I_i}\,s_{I_i}}\right\}}
{{\rm max}_i
   \left\{
      \displaystyle \sum\nolimits_j \kappa_{{I_i}{F_j}}\,s_{F_j}
   \right\}} \nn\\
&\sim & \frac{{\max_i}
   \left\{
      \displaystyle \kappa^{\rm BB}_{I_i} {1 \over
      \kappa_{I_i}} \sum\nolimits_j \kappa_{I_i F_j} {1 \over \kappa_{F_j}}
      \kappa_{F_j D} \, s_D
   \right\}}
{\max_i 
   \left\{ 
      \displaystyle \sum\nolimits_j \kappa_{I_i F_j} {1 \over
      \kappa_{F_j}} \kappa_{F_j D} \, s_D
   \right\}} \nn\\
&\sim& \frac{ \max_i 
   \left\{ 
      \displaystyle {\kappa_{I_i}^{\rm BB} \over
      \kappa_{I_i}} \sum\nolimits_j {\kappa_{I_i F_j} \over \kappa_{F_j}}
      \kappa_{F_j D}
   \right\}}
{\max_i 
   \left\{
      \displaystyle \sum\nolimits_j {\kappa_{I_i F_j} \over
      \kappa_{F_j}} \kappa_{F_j D}
   \right\}} \, ,
\label{BBregratio5}
\end{eqnarray}
where the denominators are maximized over the restricted set of
anthropic vacua $i$ (and the numerators are maximized without
restriction). The ratio of Boltzmann brains in the dominant
vacuum (vacuum $D$) to normal observers in vacua $I_i$ is given
by
\begin{eqnarray}
{\N^{\rm BB}_D\over{\N^{\rm NO}_{\{I_i\}}}} &\sim & \frac{\kappa^{\rm BB}_D
\, s_D} {\max_i \left\{ \displaystyle \sum\nolimits_j \kappa_{I_i
F_j} \, s_{F_j} \right\} } \nn \\
&\sim & \frac{\kappa^{\rm BB}_D} {{\max_i}\left\{ \displaystyle
\sum\nolimits_j {\kappa_{I_i F_j} \over \kappa_{F_j}}
\,\kappa_{F_j D} \right\}} \, ,
\label{BBregratio6}
\end{eqnarray}
and, if vacuum $D$ is anthropic, then the ratio of Boltzmann
brains in vacuum $D$ to normal observers in vacuum $D$ is given
by
\beq
{\N^{\rm BB}_D \over \N^{\rm NO}_D}
\sim  \frac{\kappa^{\rm BB}_D} {\displaystyle
\sum\nolimits_j {\kappa_{D F_j} \over \kappa_{F_j}}
\,\kappa_{F_j D} } \, .
\label{DtoD2}
\eeq

In this case our answers are complicated by the presence of many
different vacua.  We can in principle determine whether Boltzmann
brains dominate by evaluating
Eqs.~(\ref{BBregratio5})--(\ref{DtoD2}) for the correct values of
the parameters, but this gets rather complicated and
model-dependent.  The evaluation of these expressions can be
simplified significantly, however, if we make some very plausible
assumptions.

For tunneling out of the high-energy vacua $F_j$, one can expect
the transition rates into different channels to be roughly
comparable, so that $\kappa_{I_iF_j} \sim \kappa_{DF_j} \sim
\kappa_{F_j}$. That is, we assume that the branching ratios
$\kappa_{I_i F_j}/\kappa_{F_j}$ and $\kappa_{DF_j}/\kappa_{F_j}$
are roughly of order one in the sense of Eq.~(\ref{orderone}). 
These factors (or their inverses) will therefore be unimportant
in the evaluation of ${\N^{\rm BB}/{\N^{\rm NO}}}$, and may be
dropped. Furthermore, the upward transition rates from the
dominant vacuum $D$ into $F_j$ are all comparable to one another,
as can be seen by writing~\cite{LeeWeinberg}
\beq
\kappa_{F_j D}\sim e^{A_{F_j D}}e^{-S_D} \,,
\label{uprate}
\eeq
where $A_{F_j D}$ is the action of the instanton responsible for
the transition and $S_D$ is the action of the Euclideanized de
Sitter 4-sphere,
\beq
S_D = {8 \pi^2 \over H_D^2} \, .
\eeq
But generically $|A_{F_j D}|\sim 1/\rho_{F_j} $. If we assume
that
\beq
\left\vert {1 \over \rho_{F_j}} - {1 \over \rho_{F_k}}
\right\vert < 10^{14}
\label{rhobound}
\eeq
for every pair of vacua $F_j$ and $F_k$, then $\kappa_{F_j D} =
\kappa_{F_k D}$ up to a factor that can be ignored because it is
roughly of order one.  Thus, up to subleading factors, the
transition rates $\kappa_{F_j D}$ cancel out\footnote{Depending
on the range of vacua $F_j$ that are considered, the bound of
Eq.~(\ref{rhobound}) may or may not be valid.  If it is not, then
the simplification of Eq.~(\ref{leading}) below is not justified,
and the original Eq.~(\ref{BBregratio5}) has to be used.  Of
course one should remember that there was significant
arbitrariness in the choice of $10^{14}$ in the definition of
``roughly of order one.'' $10^{14}$ was chosen to accommodate the
largest estimate that we discuss in
Sec.~\ref{sec:nucleationvsdecay} for the Boltzmann brain
nucleation rate, $\Gamma_{\rm BB} \sim \exp(-10^{16})$.  In
considering the other estimates of $\Gamma_{\rm BB}$, one could
replace $10^{14}$ by a much larger number, thereby increasing the
applicability of Eq.~(\ref{rhobound}).}
in the ratio $\N^{\rm BB}_{\{I_i\}} /\N^{\rm NO}$.

Returning to Eq.~(\ref{BBregratio5}) and keeping only the leading
factors, we have
\beq
{\N^{\rm BB}_{\{I_i\}} \over{\N^{\rm NO}}} \sim
\max_i \left\{\frac{\kappa^{\rm BB}_{I_i}}{\kappa_{I_i}}\right\} \,,
\label{leading}
\eeq
where the index $i$ runs over all (non-dominant) vacua in which
Boltzmann brains can nucleate.  For the dominant vacuum, our
simplifying assumptions\footnote{The dropping of the factor
$e^{A_{F_j D}}$ is a more reliable approximation in this case
than it was in Eq.~(\ref{leading}) above.  In this case the
factor $e^{-S_D}$ does not cancel between the numerator and
denominator, so the factor $e^{A_{F_j D}}$ can be dropped if it
is unimportant compared to $e^{-S_D}$.  We of course do not know
the value of $S$ for the dominant vacuum, but for our vacuum it
is of order $10^{122}$, and it is plausible that the value for
the dominant vacuum is similar or even larger.  Thus as long as
$1/\rho_{F_j}$ is small compared to $10^{122}$, it seems safe to
drop the factor $e^{A_{F_j D}}$.} convert
Eqs.~(\ref{BBregratio6}) and (\ref{DtoD2}) into
\beq
{\N^{\rm BB}_D\over{\N^{\rm NO}}} \sim
\frac{\kappa^{\rm BB}_D}{\kappa_{\rm up}}
\sim \kappa^{\rm BB}_D \, e^{S_D} \,,
\label{BBdom}
\eeq
where $\kappa_{\rm up}\equiv \sum_j\kappa_{F_j D}$ is the upward
transition rate out of the dominant vacuum.

Thus, the conditions needed to avoid Boltzmann brain domination
are essentially the same as what we found in
Subsection~\ref{ssec:morerealistic}.  In this case, however, we
must require that in any vacuum that can support Boltzmann
brains, the Boltzmann brain nucleation rate must be less than the
decay rate of that vacuum.


\subsection{A Dominant Vacuum System}
\label{ssec:fn6}

In the next to last paragraph of Section~\ref{sec:sfcutoff}, we
described a scenario where the dominant vacuum was not the vacuum
with the smallest decay rate.  Let us now study a simple
landscape to illustrate this situation.  Consider the toy
landscape
\beq
\begin{array}{ccccccl}
&&F_j&\rightarrow& I_i&\rightarrow& B\\
&\,\,\nearrow\!\!\!\!\!\!\swarrow
&&\!\!\nwarrow\!\!\!\!\!\!\searrow&\\ A\leftarrow D_1\!\!\!\!\!\!
&&\!\!\!\!\longleftrightarrow\!\!\!\! &&\!\!\!\!D_2&\rightarrow&
A\\ &\,\,\nwarrow\!\!\!\!\!\!\searrow
&&\!\!\nearrow\!\!\!\!\!\!\swarrow&\\ &&S&\rightarrow& A\,,
\end{array}
\label{eq:fn6}
\eeq
where as in Subsection~\ref{ssec:furthergen} the vacua $I_i$ are
taken to include both anthropic vacua and vacua that support only
Boltzmann brains.  Vacua $A$ and $B$ are terminal vacua and the
$F_j$ have large, positive vacuum energies. Assume that vacuum
$S$ has the smallest total decay rate. 

We have in mind the situation in which $D_1$ and $D_2$ are nearly
degenerate, and transitions from $D_1$ to $D_2$ (and vice versa)
are rapid, even though the transition in one direction is upward.
With this in mind, we divide the decay rates of $D_1$ and $D_2$
into two parts,
\begin{eqnarray}
\kappa_{1}&=&\kappa_{21}+\kappa_{1}^{\rm out}\\
\kappa_{2}&=&\kappa_{12}+\kappa_{2}^{\rm out}\,,
\end{eqnarray}
with $\kappa_{12},\kappa_{21}\gg \kappa_{1,2}^{\rm out}$.  We
assume as in previous sections that the rates for large
upward transitions (%
$S$ to $D_1$ or $D_2$, and $D_1$ or $D_2$ to $F_j$) are extremely
small, so that we can ignore them in the calculation of $q$. The
rate equation, Eq.~(\ref{kappa-q}), then admits a solution with
$q\simeq\kappa_D$, but it also admits solutions with
\beq
q\simeq {1\over 2}\left[\kappa_1+\kappa_{2}
\pm\sqrt{(\kappa_{1}-\kappa_{2})^2+4 \kappa_{12}\kappa_{21}}\right] \,.
\label{eigq}
\eeq
Expanding the smaller root to linear order in $\kappa_{1,2}^{\rm
out}$ gives
\beq
q \simeq \alpha_1 \kappa_1^{\rm out} + \alpha_2 \kappa_2^{\rm
out}\,,
\label{altq}
\eeq
where
\beq
\alpha_1 \equiv {\kappa_{12} \over \kappa_{12}+\kappa_{21}} \ , \qquad
\alpha_2 \equiv {\kappa_{21} \over \kappa_{12}+\kappa_{21}} \,.
\label{sdef}
\eeq
In principle this value for $q$ can be smaller than $\kappa_D$,
which is the case that we wish to explore.

In this case the vacua $D_1$ and $D_2$ dominate the volume
fraction of the multiverse, even if their total decay rates
$\kappa_1$ and $\kappa_2$ are not the smallest in the landscape. 
We can therefore call the states $D_1$ and $D_2$ together a
dominant vacuum system, which we denote collectively as $D$. The
rate equation (Eq.~(\ref{kappa-q})) shows that
\beq
s_{D_1} \approx \alpha_1 \, s_D \ , \qquad s_{D_2} \approx
     \alpha_2 \, s_D
\label{sratios}
\,,
\eeq
where $s_D \equiv s_{D_1} + s_{D_2}$, and the equations hold in
the approximation that $\kappa_{1,2}^{\rm out}$ and the upward
transition rates from $D_1$ and $D_2$ can be neglected.  To see
that these vacua dominate the volume fraction, we calculate the
modified form of Eq.~(\ref{sFjratio}):
\beq
\frac{s_{F_j}}{s_D} \approx {\alpha_1 \, \kappa_{F_jD_1} +
\alpha_2 \, \kappa_{F_jD_2} \over \kappa_{F_j}} 
\, .
\eeq
Thus the volume fractions of the $F_j$, and hence also the $I_j$
and $B$ vacua, are suppressed by the very small rate for large
upward jumps from low energy vacua, namely $\kappa_{F_j D_1}$ and
$\kappa_{F_j D_2}$.  The volume fraction for $S$ depends on
$\kappa_{A D_1}$ and $\kappa_{A D_2}$, but it is maximized when
these rates are negligible, in which case it is given by
\beq
\frac{s_S}{s_D} \approx {q \over \kappa_S - q} \, .
\label{sDratio}
\eeq
This quantity can in principle be large, if $q$ is just a little
smaller than $\kappa_S$, but that would seem to be a very special
case.  Generically, we would expect that since $q$ must be
smaller than $\kappa_S$ (see Eq.~(\ref{qbound})), it would most
likely be many orders of magnitude smaller, and hence the ratio
in Eq.~(\ref{sDratio}) would be much less than one.  There is no
reason, however, to expect it to be as small as the ratios that
are suppressed by large upward jumps. For simplicity, however, we
will assume in what follows that $s_S$ can be neglected.

To calculate the ratio of Boltzmann brains to normal observers in
this toy landscape, note that Eqs.~(\ref{BBregratio5}) and
(\ref{BBregratio6}) are modified only by the substitution
\beq
  \kappa_{F_j D} \rightarrow \bar \kappa_{F_jD} \equiv \alpha_1 \,
     \kappa_{F_j D_1} + \alpha_2 \, \kappa_{F_j D_2} \, .
\eeq
Thus, the dominant vacuum transition rate is simply replaced by a
weighted average of the dominant vacuum transition rates.  If we
assume that neither of the vacua $D_1$ nor $D_2$ are anthropic,
and make the same assumptions about magnitudes used in
Subsection~\ref{ssec:furthergen}, then Eqs.~(\ref{leading}) and
(\ref{BBdom}) continue to hold as well, where we have redefined
$\kappa_{{\rm up}}$ by $\kappa_{{\rm up}} \equiv \sum_j\ \bar
\kappa_{F_jD}$.

If, however, we allow $D_1$ or $D_2$ to be anthropic, then new
questions arise.  Transitions between $D_1$ and $D_2$ are by
assumption rapid, so they copiously produce new pockets and
potentially new normal observers.  We must recall, however (as
discussed in Section \ref{sec:BBabundance}), that the properties
of a pocket universe depend on both the current vacuum and the
parent vacuum.  In this case, the unusual feature is that the
vacua within the $D$ system are nearly degenerate, and hence very
little energy is released by tunnelings within $D$. For pocket
universes created in this way, the maximum particle energy
density during reheating will be only a small fraction of the
vacuum energy density.  Such a big bang is very different from
the one that took place in our pocket, and presumably much less
likely to produce life.  We will call a vacuum in the $D$ system
``strongly anthropic'' if normal observers are produced by
tunnelings from within $D$, and ``mildly anthropic'' if normal
observers can be produced, but only by tunnelings from higher
energy vacua outside $D$.

If either of the vacua in $D$ were strongly anthropic, then the
normal observers in $D$ would dominate the normal observers in
the multiverse.  Normal observers in the vacua $I_i$ would be
less numerous by a factor proportional to the extremely small
rate $\bar \kappa_{F_jD}$ for large upward transitions .  This
situation would itself be a problem, however, similar to the
Boltzmann brain problem.  It would mean that observers like
ourselves, who arose from a hot big bang with energy densities
much higher than our vacuum energy density, would be extremely
rare in the multiverse.  We conclude that if there are any models
which give a dominant vacuum system that contains a strongly
anthropic vacuum, such models would be considered unacceptable in
the context of the scale-factor cutoff measure.

On the other hand, if the $D$ system included one or more mildly
anthropic vacua, then the situation is very similar to that
discussed in Subsections \ref{ssec:morerealistic} and
\ref{ssec:furthergen}.  In this case the normal observers in the
$D$ system would be comparable in number to the normal observers
in the vacua $I_i$, so they would have no significant effect on
the ratio of Boltzmann brains to normal observers in the
multiverse.  If any of the $D$ vacua were mildly anthropic,
however, then the stringent requirement $\kappa^{\rm BB}_D \ll
\kappa_{\rm up}$ would have to be satisfied without resort to the
simple solution $\kappa^{\rm BB}_D = 0$.

Thus, we find that the existence of a dominant vacuum system does
not change our conclusions about the abundance of Boltzmann
brains, except insofar as the Boltzmann brain nucleation
constraints that would apply to the dominant vacuum must apply to
every member of the dominant vacuum system. Probably the most
important implication of this example is that the dominant vacuum
is not necessarily the vacuum with the lowest decay rate, so the
task of identifying the dominant vacuum could be very difficult.


\subsection{General Conditions to Avoid Boltzmann Brain Domination}
\label{ssec:genconds}

In constructing general conditions to avoid Boltzmann brain
domination, we are guided by the toy landscapes discussed in the
previous subsections.  Our goal, however, is to construct
conditions that can be justified using only the general equations
of Sections~\ref{sec:sfcutoff} and~\ref{sec:BBabundance},
assuming that the landscape is irreducible, but without relying
on the properties of any particular toy landscape.  We will be
especially cautious about the treatment of the dominant vacuum
and the possibility of small upward transitions, which could be
rapid.  The behavior of the full landscape of a realistic theory
may deviate considerably from that of the simplest toy models.

To discuss the general situation, it is useful to divide vacuum
states into four classes.  We are only interested in vacua that
can support Boltzmann brains.  These can be
\begin{enumerate}
\item[(1)]  anthropic vacua for which the total dimensionless decay
rate satisfies $\kappa_i \gg q$,
\item[(2)]  non-anthropic vacua that can transition to anthropic 
vacua via unsuppressed transitions,
\item[(3)]  non-anthropic vacua that can transition to anthropic 
vacua only via suppressed transitions,
\item[(4)]  anthropic vacua for which the total dimensionless decay
rate is $\kappa_i \approx q$.
\end{enumerate}
Here $q$ is the smallest-magnitude eigenvalue of the rate
equation (see Eqs.~(\ref{rateeq})--(\ref{asymptotic})). We call a
transition ``unsuppressed'' if its branching ratio is roughly of
order one in the sense of Eq.~(\ref{orderone}). If the branching
ratio is smaller than this, it is ``suppressed.'' As before, when
calculating $\N^{\rm BB}/\N^{\rm NO}$ we assume that factors that
are roughly of order one can be ignored. Note that
Eq.~(\ref{qbound}) forbids $\kappa_i$ from being less than $q$,
so the above four cases are exhaustive. 

We first discuss conditions that are sufficient to guarantee that
Boltzmann brains will not dominate, postponing until later the
issue of what conditions are necessary.

We begin with the vacua in the first class.  Very likely all
anthropic vacua belong to this class.  For an anthropic vacuum
$i$, the Boltzmann brains produced in vacuum $i$ cannot dominate
the multiverse if they do not dominate the normal observers in
vacuum $i$, so we can begin with this comparison.  Restricting
Eq.~(\ref{BBregratio3}) to this single vacuum, we obtain
\beq
\frac{\N^{\rm BB}_i}{\N^{\rm NO}_i} \sim \frac{\kappa^{\rm BB}_i}
{\kappa_i} \, ,
\label{BBregratio7}
\eeq
a ratio that has appeared in many of the simple examples.  If
this ratio is small compared to one, then Boltzmann brains
created in vacuum $i$ are negligible. 

Let us now study a vacuum $j$ in the second class.  First note
that Eq.~(\ref{kappa-q}) implies the rigorous inequality
\beq
\kappa_i \, s_i \ge \kappa_{ij} \, s_j  \ \ \hbox{(no sum on
repeated indices)}\, ,
\label{simplebound}
\eeq
which holds for any two states $i$ and $j$.  (Intuitively,
Eq.~(\ref{simplebound}) is the statement that, in steady state,
the total rate of loss of volume fraction must exceed the input
rate from any one channel.) To simplify what follows, it will be
useful to rewrite Eq.~(\ref{simplebound}) as
\beq
(\kappa_i s_i) \ge (\kappa_j s_j) \, B_{j \rightarrow i} \, ,
\label{simplebound2}
\eeq
where $B_{j \rightarrow i} \equiv \kappa_{ij}/\kappa_j$ is the
branching ratio for the transition $j \rightarrow i$.

Suppose that we are trying to bound the Boltzmann brain
production in vacuum $j$, and we know that it can undergo
unsuppressed transitions
\beq
j \rightarrow k_1 \rightarrow \ldots \rightarrow k_n \rightarrow
i \,,
\label{sequence}
\eeq
where $i$ is an anthropic vacuum.  We begin by using
Eqs.~(\ref{Nreg}) and (\ref{Nfreak}) to express ${\N^{\rm
BB}_j}/{\N^{\rm NO}_i}$, dropping irrelevant factors as in
Eq.~(\ref{BBregratio2}), and then we can iterate the above
inequality:
\begin{eqnarray}
\frac{\N^{\rm BB}_j}{\N^{\rm NO}_i} &\sim &
{\kappa^{\rm BB}_j \, s_j \over \sum_k \kappa_{ik} \, s_k} \le 
{\kappa^{\rm BB}_j \, s_j \over \kappa_i \, s_i} \nn \\
&\le & {\kappa^{\rm BB}_j \, s_j \over (\kappa_j s_j) B_{j
\rightarrow k_1} B_{k_1 \rightarrow k_2} \cdots B_{k_n
\rightarrow i}}
\nn \\
&= & {\kappa^{BB}_j \over \kappa_j} \, {1 \over B_{j
\rightarrow k_1} B_{k_1 \rightarrow k_2} \cdots B_{k_n
\rightarrow i}} \, ,
\label{sequencebound}
\end{eqnarray}
where again there is no sum on repeated indices, and
Eq.~(\ref{kappa-q}) was used in the last step on the first line.
Each inverse branching ratio on the right of the last line is
greater than or equal to one, but by our assumptions can be
considered to be roughly of order one, and hence can be dropped.
Thus, the multiverse will avoid domination by Boltzmann brains in
vacuum $j$ if $\kappa^{\rm BB}_j/\kappa_j \ll 1$, the same
criterion found for the first class. 

The third class --- non-anthropic vacua that can only transition
to an anthropic state via at least one suppressed transition ---
presumably includes many states with very low vacuum energy
density.  The dominant vacuum of our toy landscape models
certainly belongs to this class, but we do not know of anything
that completely excludes the possibility that the dominant vacuum
might belong to the second or fourth classes.  That is, perhaps
the dominant vacuum is anthropic, or decays to an anthropic
vacuum. If there is a dominant vacuum system, as described in
Subsection~\ref{ssec:fn6}, then $\kappa_i \gg q$, and the
dominant vacua could belong to the first class, as well as to
either of classes (2) and (3).

To bound the Boltzmann brain production in this class, we
consider two possible criteria.  To formulate the first, we can
again use Eqs.~(\ref{sequence}) and~(\ref{sequencebound}), but
this time the sequence must include at least one suppressed
transition, presumably an upward jump.  Let us therefore denote
the branching ratio for this suppressed transition as $B_{\rm
up}$, noting that $B_{\rm up}$ will appear in the denominator of
Eq.~(\ref{sequencebound}). Of course, the sequence of
Eq.~(\ref{sequence}) might involve more than one suppressed
transition, but in any case the product of these very small
branching ratios in the denominator can be called $B_{\rm up}$,
and all the other factors can be taken as roughly of order one. 
Thus, a landscape containing a vacuum $j$ of the third class
avoids Boltzmann brain domination if
\beq
{\kappa^{\rm BB}_j \over B_{\rm up} \, \kappa_j} \ll 1 \, ,
\label{boundwithup}
\eeq
in agreement with the results obtained for the dominant vacua in
the toy landscape models in the previous subsections.

A few comments are in order.  First, if the only suppressed
transition is the first, then $B_{\rm up} = \kappa_{\rm
up}/\kappa_j$, and the above criterion simplifies to $\kappa^{\rm
BB}_j/\kappa_{\rm up} \ll 1$.  Second, we should keep in mind
that the sequence of Eq.~(\ref{sequence}) is presumably not
unique, so other sequences will produce other bounds.  All the
bounds will be valid, so the strongest bound is the one of
maximum interest. Finally, since the vacua under discussion are
not anthropic, a likely method for Eq.~(\ref{boundwithup}) to be
satisfied would be for $\kappa^{\rm BB}_j$ to vanish, as would
happen if the vacuum $j$ did not support the complex structures
needed to form Boltzmann brains.

The criterion above can be summarized by saying that if
$\kappa^{\rm BB}_j/(B_{\rm up} \kappa_j) \ll 1$, then the
Boltzmann brains in vacuum $j$ will be overwhelmingly outnumbered
by the normal observers living in pocket universes that form in
the decay chain starting from vacuum $j$.  We now describe a
second, alternative criterion, based on the idea that the number
of Boltzmann brains in vacuum $j$ can be compared with the number
of normal observers in vacuum $i$ if the two types of vacuum have
a common ancestor.

Denoting the common ancestor vacuum as $A$, we assume that it can
decay to an anthropic vacuum $i$ by a chain of transitions
\beq
A \rightarrow k_1 \rightarrow \ldots \rightarrow k_n \rightarrow
i \, ,
\label{sequence2}
\eeq
and also to a Boltzmann-brain-producing vacuum $j$ by a chain
\beq
A \rightarrow \ell_1 \rightarrow \ldots \rightarrow \ell_m
\rightarrow j \, .
\label{sequence3}
\eeq
From the sequence of Eq.~(\ref{sequence2}) and the bound of
Eq.~(\ref{simplebound2}), we can infer that
\beq
(\kappa_i s_i) \ge (k_A s_A) B_{A \rightarrow k_1} B_{k_1
\rightarrow k_2} \cdots B_{k_n \rightarrow i} \, .
\label{denombound}
\eeq
To make use of the sequence of Eq.~(\ref{sequence3}) we will want
a bound that goes in the opposite direction, for which will need
to require additional assumptions.  Starting with
Eq.~(\ref{kappa-q}), we first require $q \ll \kappa_i$, which is
plausible provided that vacuum $i$ is not the dominant vacuum. 
Next we look at the sum over $j$ on the right-hand side, and we
call the transition $\,j \rightarrow i\,$ ``significant'' if its
contribution to the sum is within a factor roughly of order one
of the entire sum.  (The sum over $j$ is the sum over sources for
vacuum $i$, so a transition $\,j \rightarrow i\,$ is
``significant'' if pocket universes of vacuum $j$ are a
significant source of pocket universes of vacuum $i$.) It follows
that for any significant transition $\,j \rightarrow i\,$ for
which $q \ll \kappa_i$,
\beq
(\kappa_i s_i) \le (\kappa_j s_j) Z_{\rm max} B_{j \rightarrow i}
\le (\kappa_j s_j) Z_{\rm max}
\, ,
\label{simplebound3}
\eeq
where $Z_{\rm max}$ denotes the largest number that is roughly of
order one.  By our conventions, $Z_{\rm max} = \exp(10^{14})$. 
If we assume now that all the transitions in the sequence of
Eq.~(\ref{sequence3}) are significant, and that $q$ is negligible
in each case, then
\beq
(\kappa_j s_j) \le (k_A s_A) Z_{\rm max}^{m+1} \, .
\label{numbound}
\eeq
Using the bounds from Eqs.~(\ref{denombound}) and
(\ref{numbound}), the Boltzmann brain ratio is bounded by
\begin{eqnarray}
\frac{\N^{\rm BB}_j}{\N^{\rm NO}_i} &\sim &
{\kappa^{\rm BB}_j \, s_j \over \sum_k \kappa_{ik} \, s_k} \le 
{\kappa^{\rm BB}_j \, s_j \over \kappa_i \, s_i} \nn \\
&\le& {Z_{\rm max}^{m+1} \over B_{A \rightarrow k_1} B_{k_1
\rightarrow k_2} \cdots B_{k_n \rightarrow i}}\, {\kappa^{\rm
BB}_j \over \kappa_j} \, .
\end{eqnarray}
But all the factors on the right are roughly of order one, except
that some of the branching ratios in the denominator might be
smaller, if they correspond to suppressed transitions.  If
$B_{\rm up}$ denotes the product of branching ratios for all the
suppressed transitions shown in the denominator (i.e., all
suppressed transitions in the sequence of Eq.~(\ref{sequence2})),
then the bound reduces to Eq.~(\ref{boundwithup}).\footnote{Note,
however, that the argument breaks down if the sequences in either
of Eqs.~(\ref{sequence2}) or (\ref{sequence3}) become too long. 
For the choices that we have made, a factor of $Z_{\rm max}$ is
unimportant in the calculation of $\N^{\rm BB}/\N^{\rm NO}$, but
$Z_{\rm max}^{100} = \exp(10^{16})$ can be significant.  Thus,
for our choices we can justify the dropping of ${\cal O}(100)$
factors that are roughly of order one, but not more than that.
For choices appropriate to smaller estimates of $\Gamma_{\rm
BB}$, however, the number of factors that can be dropped will be
many orders of magnitude larger.}

To summarize, the Boltzmann brains in a non-anthropic vacuum $j$
can be bounded if there is an ancestor vacuum $A$ that can decay
to $j$ through a chain of significant transitions for which $q
\ll \kappa_\ell$ for each vacuum, as in the sequence of
Eq.~(\ref{sequence3}), and if the same ancestor vacuum can decay
to an anthropic vacuum through a sequence of transitions as in
Eq.~(\ref{sequence2}).  The Boltzmann brains will never dominate
provided that $\kappa^{\rm BB}_j/(B_{\rm up} \, \kappa_j) \ll 1$,
where $B_{\rm up}$ is the product of all suppressed branching
ratios in the sequence of Eq.~(\ref{sequence2}).

Finally, the fourth class of vacua consists of anthropic vacua
$i$ with decay rate $\kappa_i \simeq q$, a class which could be
empty.  For this class Eq.~(\ref{BBregratio3}) may not be very
useful, since the quantity $(\kappa_i-q)$ in the denominator
could be very small.  Yet, as in the two previous classes, this
class can be treated by using Eq.~(\ref{sequencebound}), where in
this case the vacuum $i$ can be the same as $j$ or different,
although the case $i=j$ requires $n \ge 1$.  Again, if the
sequence contains only unsuppressed transitions, then the
multiverse avoids domination by Boltzmann brains in vacuum $i$ if
$\kappa^{\rm BB}_i/\kappa_i\ll 1$.  If upward jumps are needed to
reach an anthropic vacuum, whether it is the vacuum $i$ again or
a distinct vacuum $j$, then the Boltzmann brains in vacuum $i$
will never dominate if $\kappa^{\rm BB}_i/(B_{\rm up} \,
\kappa_i) \ll 1$.

The conditions described in the previous paragraph are very
difficult to meet, so if the fourth class is not empty, Boltzmann
brain domination is hard to avoid.  These vacua have the slowest
decay rates in the landscape, $\kappa_i
\approx q$, so it seems plausible that they have very low energy
densities, precluding the possibility of decaying to an anthropic
vacuum via unsuppressed transitions; in that case Boltzmann brain
domination can be avoided if
\beq
\kappa^{\rm BB}_i \ll B_{\rm up} \kappa_i \, .
\label{boundwithup2}
\eeq
However, as pointed out in Ref.~\cite{BFY2}, $B_{\rm up} \propto
e^{-S_D}$ (see Eq.~(\ref{uprate})) is comparable to the inverse
of the recurrence time, while in an anthropic vacuum one would
expect the Boltzmann brain nucleation rate to be much faster than
once per recurrence time.

To summarize, the domination of Boltzmann brains can be avoided
by first of all requiring that all vacuum states in the landscape
obey the relation
\beq
\frac{\kappa^{\rm BB}_j} {\kappa_j}  \ll 1 \, .
\label{bound1}
\eeq
That is, the rate of nucleation of Boltzmann brains in each
vacuum must be less than the rate of nucleation, in that same
vacuum, of bubbles of other phases. For anthropic vacua $i$ with
$\kappa_i \gg q$, this criterion is enough.  Otherwise, the
Boltzmann brains that might be produced in vacuum $j$ must be
bounded by the normal observers forming in some vacuum $i$, which
must be related to $j$ through decay chains.  Specifically, there
must be a vacuum $A$ that can decay through a chain to an
anthropic vacuum $i$, i.e. 
\beq
A \rightarrow k_1 \rightarrow \ldots \rightarrow k_n \rightarrow
i \, ,
\label{sequence4}
\eeq
where either $A=j$, or else $A$ can decay to $j$ through a
sequence
\beq
A \rightarrow \ell_1 \rightarrow \ldots \rightarrow \ell_m
\rightarrow j \, .
\label{sequence5}
\eeq
In the above sequence we insist that $\kappa_j \gg q$ and that
$\kappa_l \gg q$ for each vacuum $\ell_p$ in the chain, and that
each transition must be ``significant,'' in the sense that
pockets of type $\ell_p$ must be a significant source of pockets
of type $\ell_{p+1}$.  (More precisely, a transition from vacuum
$j$ to $i$ is ``significant'' if it contributes a fraction that
is roughly of order one to $\sum_j \kappa_{ij} s_j$ in
Eq.~(\ref{kappa-q}).) For these cases, the bound which ensures
that the Boltzmann brains in vacuum $j$ are dominated by the
normal observers in vacuum $i$ is given by
\beq
{\kappa^{\rm BB}_j \over B_{\rm up} \, \kappa_j} \ll 1 \, ,
\label{bound2}
\eeq
where $B_{\rm up}$ is the product of any suppressed branching
ratios in the sequence of Eq.~(\ref{sequence4}).  If all the
transitions in Eq.~(\ref{sequence4}) are unsuppressed, this bound
reduces to Eq.~(\ref{bound1}).  If $j$ is anthropic, the case $A
= j = i$ is allowed, provided that $n \ge 1$.

The conditions described above are sufficient to guarantee that
Boltzmann brains do not dominate over normal observers in the
multiverse, but without further assumptions there is no way to
know if they are necessary.  All of the conditions that we have
discussed are quasi-local, in the sense that they do not require
any global picture of the landscape of vacua.  For each of the
above arguments, the Boltzmann brains in one type of vacuum $j$
are bounded by the normal observers in some type of vacuum $i$
that is either the same type, or directly related to it through
decay chains.  Thus, there was no need to discuss the importance
of the vacua $j$ and $i$ compared to the rest of the landscape as
a whole.  The quasi-local nature of these conditions, however,
guarantees that they cannot be necessary to avoid the domination
by Boltzmann brains.  If two vacua $j$ and $i$ are both totally
insignificant in the multiverse, then it will always be possible
for the Boltzmann brains in vacuum $j$ to overwhelm the normal
observers in vacuum $i$, while the multiverse as a whole could
still be dominated by normal observers in other vacua.

We have so far avoided making global assumptions about the
landscape of vacua, because such assumptions are generally
hazardous.  While it may be possible to make statements that are
true for the bulk of vacua in the landscape, in this context the
statements are not useful unless they are true for {\bf all} the
vacua of the landscape.  Although the number of vacua in the
landscape, often estimated at $10^{500}$ \cite{Bousso:2004fc}, is
usually considered to be incredibly large, the number is
nonetheless roughly of order one compared to the numbers involved
in the estimates of Boltzmann brain nucleation rates and vacuum
decay rates.  Thus, if a single vacuum produces Boltzmann brains
in excess of required bounds, the Boltzmann brains from that
vacuum could easily overwhelm all the normal observers in the
multiverse. 

Recognizing that our conclusions could be faulty, we can
nonetheless adopt some reasonable assumptions to see where they
lead.  We can assume that the multiverse is sourced by either a
single dominant vacuum, or by a dominant vacuum system.  We can
further assume that every anthropic and/or
Boltzmann-brain-producing vacuum $i$ can be reached from the
dominant vacuum (or dominant vacuum system) by a single
significant upward jump, with a rate proportional to $e^{-S_D}$,
followed by some number of significant, unsuppressed transitions,
all of which have rates $\kappa_k \gg q$ and branching ratios
that are roughly of order one:
\beq
D \rightarrow k_1 \rightarrow \ldots \rightarrow k_n \rightarrow
i \, .
\label{sequence6}
\eeq
We will further assume that each non-dominant anthropic and/or
Boltzmann-brain-producing vacuum $i$ has a decay rate $\kappa_i
\gg q$, but we need not assume that all of the $\kappa_i$ are
comparable to each other.  With these assumptions, the estimate
of $\N^{\rm BB}/\N^{\rm NO}$ becomes very simple.

Applying Eq.~(\ref{kappa-q}) to the first transition of
Eq.~(\ref{sequence6}),
\beq
\kappa_{k_1} \, s_{k_1} \sim \kappa_{k_1 D} \, s_D \sim
\kappa_{\rm up} s_D \, ,
\eeq
where we use $\kappa_{\rm up}$ to denote the rate of a typical
transition $\,D \rightarrow k\,$, assuming that they are all
equal to each other up to a factor roughly of order one.  Here
$\sim$ indicates equality up to a factor that is roughly of order
one. If there is a dominant vacuum system, then $\kappa_{k_1 D}$
is replaced by $\bar \kappa_{k_1 D} \equiv \sum_\ell \alpha_\ell
\kappa_{k_1 D_\ell}$,
where the $D_\ell$ are the components of the dominant vacuum
system, and the $\alpha_\ell$ are defined by generalizing
Eqs.~(\ref{sdef}) and (\ref{sratios}).%
\footnote{In more detail, the
concept of a dominant vacuum system is relevant when there is a
set of vacua $\ell$ that can have rapid transitions within the
set, but only very slow transitions connecting these vacua to the
rest of the landscape.  As a zeroth order approximation one can
neglect all transitions connecting these vacua to the rest of the
landscape, and assume that $\kappa_\ell \gg q$, so
Eq.~(\ref{kappa-q}) takes the form
\[
  \kappa_\ell \, s_\ell = \sum_{\ell'} B_{\ell \ell'} \,
     \kappa_{\ell'} \, s_{\ell'} \, .
\]
Here $B_{\ell \ell'} \equiv \kappa_{\ell \ell'} /
\kappa_{\ell'}$
is the branching ratio within this restricted subspace, where
$\kappa_{\ell} = \sum\nolimits_{\ell'} \, \kappa_{\ell' \ell}$ is
summed only within the dominant vacuum system, so $\sum_\ell
B_{\ell \ell'} = 1$ for all $\ell'$.  $B_{\ell \ell'}$ is
nonnegative, and if we assume also that it is irreducible, then
the Perron-Frobenius theorem guarantees that it has a
nondegenerate eigenvector $v_\ell$ of eigenvalue 1, with positive
components.  From the above equation $\kappa_\ell \, s_\ell
\propto v_\ell$, and then
\[
\alpha_\ell = {s_\ell \over \sum_{\ell'} s_{\ell'}} = {v_\ell
\over \displaystyle \kappa_\ell \sum_{\ell'} {v_{\ell'} \over
\kappa_{\ell'}}} \, .
\]
}
Applying Eq.~(\ref{kappa-q}) to the next transition, $\,k_1
\rightarrow k_2\,$ we find
\beq
\kappa_{k_2} \, s_{k_2} = B_{k_1 \rightarrow k_2} \kappa_{k_1}
s_{k_1} + \ldots \sim \kappa_{k_1} \, s_{k_1} \, ,
\eeq
where we have used the fact that $B_{k_1 \rightarrow k_2}$ is
roughly of order one, and that the transition is significant. 
Iterating, we have
\beq
\kappa_i \, s_i \sim \kappa_{k_n} \, s_{k_n} \sim \kappa_{\rm up}
\label{kappa-up}
\, s_D \, .
\eeq
Since the expression on the right is independent of $i$, we
conclude that under these assumptions any two non-dominant
anthropic and/or Boltzmann-brain-producing vacua $i$ and $j$ have
equal values of $\kappa s$, up to a factor that is roughly of
order one:
\beq
\kappa_j s_j \sim \kappa_i s_i \ . 
\label{kappa-s}
\eeq
Using Eq.~(\ref{Nreg}) and assuming as always that $n^{\rm
NO}_{ik}$ is roughly of order one, Eq.~(\ref{kappa-s}) implies
that any two non-dominant anthropic vacua $i$ and $j$ have
comparable numbers of ordinary observers, up to a factor that is
roughly of order one:
\beq
\N^{\rm NO}_j \sim \N^{\rm NO}_i \ . 
\eeq

The dominant vacuum could conceivably be anthropic, but we begin
by considering the case in which it is not.  In that case all
anthropic vacua are equivalent, so the Boltzmann brains produced
in any vacuum $j$ will either dominate the multiverse or not
depending on whether they dominate the normal observers in an
arbitrary anthropic vacuum $i$. Combining Eqs.~(\ref{Nreg}),
(\ref{Nfreak}), (\ref{kappa-q}), and (\ref{kappa-s}), and
omitting irrelevant factors, we find that for any non-dominant
vacuum $j$
\beq
{\N^{\rm BB}_j \over \N^{\rm NO}_i} \sim {\kappa^{\rm BB}_j s_j
\over \sum_k \kappa_{ik} \, s_k} \sim {\kappa^{\rm BB}_j s_j\over
\kappa_i s_i} \sim {\kappa^{\rm BB}_j \over \kappa_j } \, .
\eeq
Thus, given the assumptions described above, for any non-dominant
vacuum $j$ the necessary and sufficient condition to avoid the
domination of the multiverse by Boltzmann brains in vacuum $j$ is
given by
\beq
{\kappa^{\rm BB}_j \over \kappa_j} \ll 1 \, .
\eeq

For Boltzmann brains formed in the dominant vacuum, we can again
find out if they dominate the multiverse by determining whether
they dominate the normal observers in an arbitrary anthropic
vacuum $i$.  Repeating the above analysis for vacuum $D$ instead
of vacuum $j$, using Eq.~(\ref{kappa-up}) to relate $s_i$ to
$s_D$, we have
\beq
{\N^{\rm BB}_D \over \N^{\rm NO}_i} \sim {\kappa^{\rm BB}_D s_D
\over \sum_k \kappa_{ik} \, s_k} \sim {\kappa^{\rm BB}_D s_D\over
\kappa_i s_i} \sim {\kappa^{\rm BB}_D \over \kappa_{\rm up}} \, .
\eeq
Thus, for a single dominant vacuum $D$ or a dominant vacuum
system with members $D_i$, the necessary and sufficient
conditions to avoid the domination of the multiverse by these
Boltzmann brains is given by
\beq
{\kappa^{\rm BB}_D \over \kappa_{\rm up}} \ll 1 \ \hbox{ or } \
{\kappa^{\rm BB}_{D_i} \over \kappa_{\rm up}} \ll 1 \, .
\label{Dbound}
\eeq
As discussed after Eq.~(\ref{boundwithup2}), probably the only
way to satisfy this condition is to require that $\kappa^{\rm
BB}_D = 0$.

If the dominant vacuum is anthropic, then the conclusions are
essentially the same, but the logic is more involved.  For the
case of a dominant vacuum system, we distinguish between the
possibility of vacua being ``strongly'' or ``mildly'' anthropic,
as discussed in Subsection
\ref{ssec:fn6}.  ``Strongly anthropic'' means that normal
observers are formed by tunneling within the dominant vacuum
system $D$, while ``mildly anthropic'' implies that normal
observers are formed by tunneling, but only from outside $D$. 
Any model that leads to a strongly anthropic dominant vacuum
would be unacceptable, because almost all observers would live in
pockets with a maximum reheat energy density that is small
compared to the vacuum energy density.  With a single anthropic
dominant vacuum, or with one or more mildly anthropic vacua
within a dominant vacuum system, the normal observers in the
dominant vacuum would be comparable in number (up to factors
roughly of order one) to those in other anthropic vacua, so they
would have no significant effect on the ratio of Boltzmann brains
to normal observers in the multiverse.  An anthropic vacuum would
also produce Boltzmann brains, however, so Eq.~(\ref{Dbound})
would have to somehow be satisfied for $\kappa^{\rm BB}_D \not =
0$.


\section{Boltzmann Brain Nucleation and Vacuum Decay Rates}
\label{sec:nucleationvsdecay}


\subsection{Boltzmann Brain Nucleation Rate}
\label{sec:BBnucleation}

Boltzmann brains emerge from the vacuum as large quantum
fluctuations.  In particular, they can be modeled as localized
fluctuations of some mass $M$, in the thermal bath of a de Sitter
vacuum with temperature $T_{\rm dS} = H_\Lambda/2 \pi$ \cite{GH}. 
The Boltzmann brain nucleation rate is then roughly estimated by
the Boltzmann suppression factor~\cite{Page1,BF06},
\beq
\Gamma_{\rm BB} \sim e^{-M/T_{\rm dS}} \,,
\label{Gamma1}
\eeq
where our goal is to estimate only the exponent, not the
prefactor.  Eq.~(\ref{Gamma1}) gives an estimate for the
nucleation rate of a Boltzmann brain of mass $M$ in any
particular quantum state, but we will normally describe the
Boltzmann brain macroscopically.  Thus $\Gamma_{\rm BB}$ should
be multiplied by the number of microstates $e^{S_{\rm BB}}$
corresponding to the macroscopic description, where $S_{\rm BB}$
is the entropy of the Boltzmann brain.  Thus we expect
\beq
\Gamma_{\rm BB}\sim e^{-M/T_{\rm dS}}e^{S_{\rm BB}} 
= e^{-F/T_{\rm dS}} \,,
\label{Gamma2}
\eeq
where $F = M - T_{\rm dS} \, S_{\rm BB}$ is the free energy of
the Boltzmann brain.

Eq.~(\ref{Gamma2}) should be accurate as long as the de Sitter
temperature is well-defined, which will be the case as long as
the Schwarzschild horizon is small compared to the de Sitter
horizon radius.  Furthermore, we shall neglect the effect of the
gravitational potential energy of de Sitter space on the
Boltzmann brain, which requires that the Boltzmann brain be small
compared to the de Sitter horizon.  Thus we assume
\beq
M/4\pi < R \ll H^{-1}_{\Lambda}\,,
\label{smallMass}
\eeq  
where the first inequality assumes that Boltzmann brains cannot
be black holes.  The general situation, which allows for $M\sim
R\sim H_\Lambda^{-1}$, will be discussed in Appendix~\ref{app:1}
and in Ref.~\cite{BBfuture}.

While the nucleation rate is proportional to $e^{S_{\rm BB}}$,
this factor is negligible for any Boltzmann brain made of atoms
like those in our universe.  The entropy of such atoms is bounded
by
\beq
S\lesssim 3 M/m_n \,,
\label{S}
\eeq 
where $m_n$ is the nucleon mass.  Indeed, the actual value of
$S_{\rm BB}$ is much smaller than this upper bound because of the
complex organization of the Boltzmann brain.  Meanwhile, to
prevent the Boltzmann brain from being destroyed by pair
production, we require that $T_{\rm dS}\ll m_n$. Thus, for these
Boltzmann brains the entropy factor $e^{S_{\rm BB}}$ is
irrelevant compared to the Boltzmann suppression factor. 

To estimate the nucleation rate for Boltzmann brains, we need at
least a crude description of what constitutes a Boltzmann brain. 
There are many possibilities.  We argued in the introduction to
this paper that a theory that predicts the domination of
Boltzmann brains over normal observers would be overwhelmingly
disfavored by our continued observation of an orderly world, in
which the events that we observe have a logical relationship to
the events that we remember.  In making this argument, we
considered a class of Boltzmann brains that share exactly the
memories and thought processes of a particular normal observer at
some chosen instant.  For these purposes the memory of the
Boltzmann brain can consist of random bits that just happen to
match those of the normal observer, so there are no requirements
on the history of the Boltzmann brain.  Furthermore, the
Boltzmann brain need only survive long enough to register one
observation after the chosen instant, so it is not required to
live for more than about a second.  We will refer to Boltzmann
brains that meet these requirements as minimal Boltzmann brains. 

While an overabundance of minimal Boltzmann brains is enough to
cause a theory to be discarded, we nonetheless find it
interesting to discuss a wide range of Boltzmann brain
possibilities.  We will start with very large Boltzmann brains,
discussing the minimal Boltzmann brains last.

We first consider Boltzmann brains much like us, who evolved in
stellar systems like ours, in vacua with low-energy particle
physics like ours, but allowing for a de Sitter Hubble radius as
small as a few astronomical units or so.  These Boltzmann brains
evolved in their stellar systems on a time scale similar to the
evolution of life on Earth, so they are in every way like us,
except that, when they perform cosmological observations, they
find themselves in an empty, vacuum-dominated universe.  These
``Boltzmann solar systems'' nucleate at a rate of roughly
\beq 
\Gamma_{\rm BB} \sim \exp(-10^{85}) \,,
\eeq 
where we have set $M\sim 10^{30}$~kg and $H_\Lambda^{-1}=(2\pi
T_{\rm dS})^{-1} \sim 10^{12}$~m.  This nucleation rate is
fantastically small; we found it, however, by considering the
extravagant possibility of nucleating an entire Boltzmann solar
system.

Next, we can consider the nucleation of an isolated brain, with a
physical construction that is roughly similar to our own brains. 
If we take $M\sim 1$~kg and $H_\Lambda^{-1}=(2\pi T_{\rm
dS})^{-1} \sim 1$~m, then the corresponding Boltzmann brain
nucleation rate is
\beq
\Gamma_{\rm BB}\sim \exp(-10^{43}) \,.
\label{1045}
\eeq
If the construction of the brain is similar to ours, however,
then it could not function if the tidal forces resulted in a
relative acceleration from one end to the other that is much
greater than the gravitational acceleration $g$ on the surface of
the Earth.  This requires $H_\Lambda^{-1}\gtrsim 10^8$~m, giving
a Boltzmann brain nucleation rate
\beq
\Gamma_{\rm BB}\sim \exp(-10^{51}) \,.
\label{1051}
\eeq

Until now, we have concentrated on Boltzmann brains that are very
similar to human brains.  However a common assumption in the
philosophy of mind is that of substrate-independence. Therefore,
pressing onward, we study the possibility that a Boltzmann brain
can be any device capable of emulating the thoughts of a human
brain.  In other words, we treat the brain essentially as a
highly sophisticated computer, with logical operations that can
be duplicated by many different systems of
hardware.\footnote{Note that the validity of the assumption of
substrate-independence of mind is not entirely self-evident. 
Some of us are skeptical of identifying human consciousness with
operations of a generic substrate-independent computer, but
accept it as a working hypothesis for the purpose of this paper.}

With this in mind, from here out we drop the assumption that
Boltzmann brains are made of the same materials as human brains. 
Instead, we attempt to find an upper bound on the probability of
creation of a more generalized computing device, specified by its
information content $I_{\rm BB}$, which is taken to be comparable
to the information content of a human brain.

To clarify the meaning of information content, we can model an
information storage device as a system with $N$ possible
microstates.  $S_{\rm max}=\ln N$ is then the maximum entropy
that the system can have, the entropy corresponding to the state
of complete uncertainty of microstate.  To store $B$ bits of
information in the device, we can imagine a simple model in which
$2^B$ distinguishable macroscopic states of the system are
specified, each of which will be used to represent one assignment
of the bits.  Each macroscopic state can be modeled as a mixture
of $N/2^B$ microstates, and hence has entropy $S = \ln (N/2^B) =
S_{\rm max} - B \ln 2$.  Motivated by this simple model, one
defines the information content of any macroscopic state of
entropy $S$ as the difference between $S_{\rm max}$ and $S$,
where $S_{\rm max}$ is the maximum entropy that the device can
attain.  Applying this definition to a Boltzmann brain, we write
\beq
I_{\rm BB} = S_{\rm BB,max} - S_{\rm BB} \, ,
\label{Idefine}
\eeq
where $I_{\rm BB}/\!\ln 2$ is the information content measured in
bits.

As discussed in Ref.~\cite{Lloyd}, the only known
substrate-independent limit on the storage of information is the
Bekenstein bound. It states that, for an asymptotically flat
background, the entropy of any physical system of size $R$ and
energy $M$ is bounded by\footnote{In an earlier version of this
paper we stated an incorrect form of this bound, and from it
derived some incorrect conclusions, such as the statement that
the largest Boltzmann brain nucleation rate $\Gamma_{\rm BB}$
consistent with the Bekenstein bound is attained only when the
radius $R$ approaches the Schwarzschild radius $R_{\rm Sch}$. 
This in turn led to the conclusion that the maximum rate allowed
by the Bekenstein bound is $e^{-2 I_{\rm BB}}$, which can be
achieved only if $M^2 = I_{\rm BB}/(9 \pi G)$ and $H_\Lambda^2 =
\pi/(3 G I_{\rm BB})$.  While these relations hold in the regime
we considered, they are not necessary in the general case.  With
the corrected bound, we find that the maximum nucleation rate is
independent of $R/R_{\rm Sch}$ if $R \ll H_\Lambda$ (see
Eq.~(\ref{Gamma3})), and otherwise grows with $R/R_{\rm Sch}$
(see Appendix~\ref{app:1}).  However, once one is forced to
consider values of $R \gg R_{\rm Sch}$, then other issues become
relevant.  How can the system be stabilized against the de Sitter
expansion? Can the Bekenstein bound really be saturated for a
system with large entropy, especially if it is dilute? In this
version of the paper we have added a discussion of these issues. 
We thank R.~Bousso, B.~Freivogel, and I.~Yang for pointing out
the error in our earlier statement of the Bekenstein bound.}
\beq
S \leq S_{\rm Bek} \equiv 2 \pi M R \,.
\label{BekensteinB}
\eeq
One can use this bound in de Sitter space as well if the size of
the system is sufficiently small, $R \ll H_{\Lambda}^{-1}$, so
that the system does not ``know'' about the horizon.  A possible
generalization of the Bekenstein bound for $R={\mathcal
O}(H_{\Lambda}^{-1})$ was proposed in Ref.~\cite{Bousso2000}; we
will study this and other possibilities in Appendix~\ref{app:1}
and in Ref.~\cite{BBfuture}.  To begin, however, we will discuss
the simplest case, $R\ll H_{\Lambda}^{-1}$, so that we can focus
on the most important issues before dealing with the complexities
of more general results.

Using Eq.~(\ref{Idefine}), the Boltzmann brain nucleation rate of
Eq.~(\ref{Gamma2}) can be rewritten as
\beq
\Gamma_{\rm BB} \sim \exp \left( -\frac{2\pi M}{H_{\Lambda}} +
S_{\rm BB,max} - I_{\rm BB} \right) \, ,
\eeq
which is clearly maximized by choosing $M$ as small as possible. 
The Bekenstein bound, however, implies that $S_{\rm BB,max} \le
S_{\rm Bek}$ and therefore $M \ge S_{\rm BB,max}/(2 \pi R)$. 
Thus
\beq
\Gamma_{\rm BB} \le \exp \left( -\frac{S_{\rm BB,max}}{R H_{\Lambda}} +
S_{\rm BB,max} - I_{\rm BB} \right) \, .
\eeq
Since $R < H_{\Lambda}^{-1}$, the expression above is maximized
by taking $S_{\rm BB,max}$ equal to its smallest possible value,
which is $I_{\rm BB}$.  Finally, we have
\beq
\Gamma_{\rm BB} \leq \exp \left( -\frac{I_{\rm BB}}{RH_{\Lambda}}
\right) \, . \label{Gamma3}
\eeq

Thus, the Boltzmann brain production rate is maximized if the
Boltzmann brain saturates the Bekenstein bound, with $I_{\rm
BB}=S_{\rm BB,max} = 2\pi MR$. Simultaneously, we should make
$RH_{\Lambda}$ as large as possible, which means taking our
assumption $R\ll H_{\Lambda}^{-1}$ to the boundary of its
validity. Thus we write the Boltzmann brain production rate
\beq
\Gamma_{\rm BB} \leq e^{- a I_{\rm BB}}\,,
\label{BBlimit}
\eeq
where $a \equiv (RH_{\Lambda})^{-1}$, the value of which is of
order a few.  In Appendix~\ref{app:1} we explore the case in
which the Schwarzschild radius, the Boltzmann brain radius, and
the de Sitter horizon radius are all about equal, in which case
Eq.~(\ref{BBlimit}) holds with $a=2$.

The bound of Eq.~(\ref{BBlimit}) can be compared to the estimate
of the Boltzmann brain production rate, $\Gamma_{\rm BB}\sim
e^{-S_{\rm BB}}$, which follows from Eq.~(2.13) of Freivogel and
Lippert, in Ref.~\cite{FreivogelLippert}.  The authors of
Ref.~\cite{FreivogelLippert} explained that by $S_{\rm BB}$ they
mean not the entropy, but the number of degrees of freedom, which
is roughly equal to the number of particles in a Boltzmann brain. 
This estimate appears similar to our result, if one equates
$S_{\rm BB}$ to $I_{\rm BB}$, or to a few times $I_{\rm BB}$.
Freivogel and Lippert describe this relation as a lower bound on
the nucleation rate for Boltzmann brains, commenting that it can
be used as an estimate of the nucleation rate for vacua with
``reasonably cooperative particle physics.'' Here we will explore
in some detail the question of whether this bound can be used as
an estimate of the nucleation rate.  While we will not settle
this issue here, we will discuss evidence that
Eq.~(\ref{BBlimit}) is a valid estimate for at most a small
fraction of the vacua of the landscape, and possibly none at all.

So far, the conditions to reach the upper bound in
Eq.~(\ref{BBlimit}) are $R = (aH_\Lambda)^{-1} \sim {\cal
O}(H_\Lambda^{-1})$ and $I_{\rm BB} = S_{\rm max,BB} = S_{\rm
Bek}$. However these are not enough to ensure that a Boltzmann
brain of size $R\sim H_\Lambda^{-1}$ is stable and can actually
compute.  Indeed, the time required for communication between two
parts of a Boltzmann brain separated by a distance ${\cal
O}(H_\Lambda^{-1})$ is at least comparable to the Hubble time. 
If the Boltzmann brain can be stretched by cosmological
expansion, then after just a few operations the different parts
will no longer be able to communicate.  Therefore we need a
stabilization mechanism by which the brain is protected against
expansion.

A potential mechanism to protect the Boltzmann brain against de
Sitter expansion is the self-gravity of the brain.  A simple
example is a black hole, which does not expand when the universe
expands.  It seems
unlikely that black holes can think,\footnote{The possibility of
a black hole computer is not excluded, however, and has been
considered in Ref.~\cite{Lloyd}.  Nonetheless, if black holes can
compute, our conclusions would not be changed, provided that the
Bekenstein bound can be saturated for the near-black hole
computers that we discuss.  At this level of approximation, there
would be no significant difference between a black hole computer
and a near-black hole computer.} 
but one can consider objects of mass approaching that of a black
hole with radius $R$. This, together with our goal to keep $R$ as
close as possible to $H_\Lambda^{-1}$, leads to the following
condition:
\beq
M \sim R \sim H_\Lambda^{-1}\,.
\label{BHconstraint}
\eeq
If the Bekenstein bound is saturated, this leads to the following
relations between $I_{\rm BB}$, $H_\Lambda$, and $M$:
\beq
I_{\rm BB} \sim MR \sim MH_\Lambda^{-1} \sim H_\Lambda^{-2} \,.
\label{BHconstraint2}
\eeq

A second potential mechanism of Boltzmann brain stabilization is
to surround it by a domain wall with a surface tension $\sigma$,
which would provide pressure preventing the exponential expansion
of the brain.  An investigation of this situation reveals that
one cannot saturate the Bekenstein bound using this mechanism
unless there is a specific relation between $I_{\rm BB}$,
$H_\Lambda$, and $\sigma$~\cite{BBfuture}:
\beq
\sigma \sim  I_{\rm BB}\, H_\Lambda^{3} \,.
\label{ShellConstraint2}
\eeq
If $\sigma$ is less than this magnitude, it cannot prevent the
expansion, while a larger $\sigma$ increases the mass and
therefore prevents saturation of the Bekenstein bound.

Regardless of the details leading to Eqs.~(\ref{BHconstraint2})
and (\ref{ShellConstraint2}), the important point is that both of
them lead to constraints on the vacuum hosting the Boltzmann
brain.

For example, the Boltzmann brain stabilized by gravitational
attraction can be produced at a rate approaching $e^{- a I_{\rm
BB}}$ only if $I_{\rm BB} \sim H_{\Lambda}^{-2}$.  For a given
value of $I_{\rm BB}$, say $I_{\rm BB}\sim 10^{16}$ (see the
discussion below), this result applies only to vacua with a
particular vacuum energy, $\Lambda\sim 10^{-16}$.  Similarly,
according to Eq.~(\ref{ShellConstraint2}), for Boltzmann brains
with $I_{\rm BB}\sim 10^{16}$ contained inside a domain wall in a
vacuum with $\Lambda \sim 10^{-120}$, the Bekenstein bound on
$\Gamma_{\rm BB}$ cannot be reached unless the tension of the
domain wall is incredibly small, $\sigma \sim 10^{-164}$.  Thus,
the maximal Boltzmann brain production rate $\sim e^{- a I_{\rm
BB}}$ saturating the Bekenstein bound cannot be reached unless
Boltzmann brains are produced on a narrow hypersurface in the
landscape.

This conclusion by itself does not eliminate the danger of a
rapid Boltzmann brain production rate, $\Gamma_{\rm BB}\sim e^{-
a I_{\rm BB}}$.  Given the vast number of vacua in the landscape,
it seems plausible that this bound could actually be met. If this
is the case, Eq.~(\ref{BBlimit}) offers a stunning increase over
previous estimates of $\Gamma_{\rm BB}$.

Setting aside the issue of Boltzmann brain stability, one can
also question the assumption of Bekenstein bound saturation that
is necessary to achieve the rather high nucleation rate that is
indicated by Eq.~(\ref{BBlimit}). Of course black holes saturate
this bound, but we assume that a black hole cannot think.  Even
if a black hole can think, it would still be an open question
whether this information processing could make use of a
substantial fraction of the degrees of freedom associated with
the black hole entropy. A variety of other physical systems are
considered in Ref.~\cite{Bekenstein:1984}, where the validity of
$S_{\rm max}(E) \leq 2\pi ER$ is studied as a function of energy
$E$.  In all cases, the bound is saturated in a limit where
$S_{\rm max} = {\cal O}(1)$. Meanwhile, as we shall argue below,
the required value of $S_{\rm max}$ should be greater than
$10^{16}$.

The present authors are aware of only one example of a physical
system that may saturate the Bekenstein bound and at the same
time store sufficient information $I$ to emulate a human brain. 
This may happen if the total number of particle species with mass
smaller than $H_\Lambda$ is greater than $I_{\rm BB}\gtrsim
10^{16}$.  No realistic examples of such theories are known to
us, although some authors have speculated about similar
possibilities~\cite{Dvali}.

If Boltzmann brains cannot saturate the Bekenstein bound, they
will be more massive than indicated in Eq.~(\ref{Gamma3}), and
their rate of production will be smaller than $e^{-aI_{\rm BB}}$.

To put another possible bound on the probability of Boltzmann
brain production, let us analyze a simple model based on an ideal
gas of massless particles.  Dropping all numerical factors, we
consider a box of size $R$ filled with a gas with maximum entropy
$S_{\rm max} = (RT)^3$ and energy $E = R^3T^4 = S_{\rm
max}^{4/3}/R$, where $T$ is the temperature and we assume there
is not an enormous number of particle species.  The probability
of its creation can be estimated as follows:
\beq
\Gamma_{\rm BB} \sim e^{-E/H_\Lambda}\,e^{S_{\rm BB}} 
\sim \exp\left( -\frac{S_{\rm max}^{4/3}}{H_\Lambda R} \right) ,
\label{pre-fourthirds}
\eeq
where we have neglected the Boltzmann brain entropy factor, since
$S_{\rm BB} \leq S_{\rm max} \ll S_{\rm max}^{4/3}$.  This
probability is maximized by taking $R \sim H_\Lambda^{-1}$, which
yields
\beq
\Gamma_{\rm BB} \lesssim e^{-S_{\rm max}^{4/3}} \ .
\eeq
In case the full information capacity of the gas is used, one can
also write
\beq
\Gamma_{\rm BB} \lesssim e^{-I_{\rm BB}^{4/3}} \ .
\label{fourthirds}
\eeq
For $I_{\rm BB} \gg 1$, this estimate leads to a much stronger
suppression of Boltzmann brain production as compared to our
previous estimate, Eq.~(\ref{BBlimit}). 

Of course, such a hot gas of massless particles cannot think ---
indeed it is not stable in the sense outlined below
Eq.~(\ref{BBlimit}) --- so we must add more parts to this
construction.  Yet it seems likely that this will only decrease
the Boltzmann brain production rate.  As a partial test of this
conjecture, one can easily check that if instead of a gas of
massless particles we consider a gas of massive particles, the
resulting suppression of Boltzmann brain production will be
stronger.  Therefore in our subsequent estimates we shall assume
that Eq.~(\ref{fourthirds}) represents our next ``line of
defense'' against the possibility of Boltzmann brain domination,
after the one given by Eq.~(\ref{BBlimit}).  One should note that
this is a rather delicate issue; see for example a discussion of
several possibilities to approach the Bekenstein bound in
Ref.~\cite{Bousso:2010pm}.  A more detailed discussion of this
issue will be provided in Ref.~\cite{BBfuture}.

Having related $\Gamma_{\rm BB}$ to the information content
$I_{\rm BB}$ of the brain, we now need to estimate $I_{\rm BB}$. 
How much information storage must a computer have to be able to
perform all the functions of the human brain? Since no one can
write a computer program that comes close to imitating a human
brain, this is not an easy question to answer.

One way to proceed is to examine the human brain, with the goal
of estimating its capacities based on its biological structure.
The human brain contains $\sim\nobreak 10^{14}$ synapses that may
in principle connect to any of $\sim\nobreak 10^{11}$
neurons~\cite{Pakkenberg}, suggesting that its information
content\footnote{Note that the specification of one out of
$10^{11}$ neurons requires $\log_2\left(10^{11}\right) = 36.5$
bits.} 
might be roughly $I_{\rm BB} \sim 10^{15}$--$10^{16}$. (We
are assuming here that the logical functions of the brain depend
on the connections among neurons, and not for example on their
precise locations, cellular structures, or other information that
might be necessary to actually construct a brain.) A minimal
Boltzmann brain is only required to simulate the workings of a
real brain for about a second, but with neurons firing typically
at 10 to 100 times a second, it is plausible that a substantial
fraction of the brain is needed even for only one second of
activity.  Of course the actual number of required bits might be
somewhat less.

An alternative approach is to try to determine how much
information the brain processes, even if one does not understand
much about what the processing involves.

In Ref.~\cite{Landauer1}, Landauer attempted to estimate the
total content of a person's long-term memory, using a variety of
experiments.  He concluded that a person remembers only about 2
bits/second, for a lifetime total in the vicinity of $10^9$ bits. 
In a subsequent paper~\cite{Landauer2}, however, he emphatically
denied that this number is relevant to the information
requirements of a ``real or theoretical cognitive processor,''
because such a device ``would have so much more to do than simply
record new information.''

Besides long-term memory, one might be interested in the total
amount of information a person receives but does not memorize. A
substantial part of this information is visual; it can be
estimated by the information stored on high definition DVDs
watched continuously on several monitors over the span of a
hundred years. The total information received would be about
$10^{16}$ bits.

Since this number is similar to the number obtained above by
counting synapses, it is probably as good an estimate as we can
make for a minimal Boltzmann brain.  If the Bekenstein bound can
be saturated, then the estimated Boltzmann brain nucleation rate
for the most favorable vacua in the landscape would be given by
Eq.~(\ref{BBlimit}):
\beq
\Gamma_{\rm BB} \lesssim e^{-10^{16}} \, .
\label{computer}
\eeq
If, however, the Bekenstein bound cannot be reached for systems
with $I_{\rm BB} \gg 1$, then it might be more accurate to use
instead the ideal gas model of Eq.~(\ref{fourthirds}), yielding
\beq
\Gamma_{\rm BB} \lesssim e^{-10^{21}} \, .
\label{1021}
\eeq

Obviously, there are many uncertainties involved in the numerical
estimates of the required value of $I_{\rm BB}$.  Our estimate
$I_{\rm BB}\sim 10^{16}$ concerns the information stored in the
human brain that appears to be relevant for cognition.  It
certainly does not include all the information that would be
needed to physically construct a human brain, and it therefore
does not allow for the information that might be needed to
physically construct a device that could emulate the human brain.
\footnote{\label{wiring}That is, the actual construction of a
brain-like device would presumably require large amounts of
information that are not part of the schematic ``circuit
diagram'' of the brain.  Thus there may be some significance to
the fact that a billion years of evolution on Earth has not
produced a human brain with fewer than about $10^{27}$ particles,
and hence of order $10^{27}$ units of entropy.  In counting the
information in the synapses, for example, we counted only the
information needed to specify which neurons are connected to
which, but nothing about the actual path of the axons and
dendrites that complete the connections.  These are nothing like
nearest-neighbor couplings, but instead axons from a single
neuron can traverse large fractions of the brain, resulting in an
extremely intertwined network \cite{DyanAbbott}.  To specify even
the topology of these connections, still ignoring the precise
locations, could involve much more than $10^{16}$ bits.  For
example, the synaptic ``wiring'' that connects the neurons will
in many cases form closed loops. A specification of the
connections would presumably require a topological winding number
for every pair of closed loops in the network.  The number of
bits required to specify these winding numbers would be
proportional to the square of the number of closed loops, which
would be proportional to the square of the number of synapses. 
Thus, the structural information could be something like $I_{\rm
struct} \sim b \times 10^{28}$, where $b$ is a proportionality
constant that is probably a few orders of magnitude less than 1. 
In estimating the resulting suppression of the nucleation rate,
there is one further complication: since structural information
of this sort presumably has no influence on brain function, these
choices would contribute to the multiplicity of Boltzmann brain
microstates, thereby multiplying the nucleation rate by
$e^{I_{\rm struct}}$.  There would still be a net suppression,
however, with Eq.~(\ref{BBlimit}) leading to $\Gamma_{\rm BB}
\propto e^{-(a-1) I_{\rm struct}}$, where $a$ is generically
greater than 1.  See Appendix~\ref{app:1} for further discussion
of the value of $a$.}
It is also possible that extra mass might be required for the
mechanical structure of the emulator, to provide the analogues of
a computer's wires, insulation, cooling systems, etc. On the
other hand, it is conceivable that a Boltzmann brain can be
relevant even if it has fewer capabilities than what we called
the minimal Boltzmann brain.  In particular, if our main
requirement is that the Boltzmann brain is to have the same
``perceptions'' as a human brain for just one second, then one
may argue that this can be achieved using much less than
$10^{14}$ synapses. And if one decreases the required time to a
much smaller value required for a single computation to be
performed by a human brain, the required amount of information
stored in a Boltzmann brain may become many orders of magnitude
smaller than $10^{16}$.
 
We find that regardless of how one estimates the information in a
human brain, if Boltzmann brains can be constructed so as to come
near the limit of Eq.~(\ref{BBlimit}), their nucleation rate
would provide stringent requirements on vacuum decay rates in the
landscape.  On the other hand, if no such physical construction
exists, we are left with the less dangerous bound of
Eq.~(\ref{fourthirds}), perhaps even further softened by the
speculations described in Footnote \ref{wiring}. Note that none
of these bounds is based upon a realistic model of a Boltzmann
brain.  For example, the nucleation of an actual human brain is
estimated at the vastly smaller rate of Eq.~(\ref{1051}).  The
conclusions of this paragraph apply to the causal patch
measures~\cite{diamond,censor} as well as the scale-factor cutoff
measure.

In Section~\ref{sec:BBabundance} we discussed the possibility of
Boltzmann brain production during reheating, stating that this
process would not be a danger.  We postponed the numerical
discussion, however, so we now return to that issue.  According
to Eq.~(\ref{Domreheat}), the multiverse will be safe from
Boltzmann brains formed during reheating provided that
\beq
\Gamma^{\rm BB}_{{\rm reheat},ik} \, \Delta \tau^{\rm BB}_{{\rm
reheat},ik} \ll n^{\rm NO}_{ik}
\label{reheatbound}
\eeq
holds for every pair of vacua $i$ and $k$, where $\Gamma^{\rm
BB}_{{\rm reheat},ik}$ is the peak Boltzmann brain nucleation
rate in a pocket of vacuum $i$ that forms in a parent vacuum of
type $k$, $\Delta \tau^{\rm BB}_{{\rm reheat},ik}$ is the proper
time available for such nucleation, and $n^{\rm NO}_{ik}$ is the
volume density of normal observers in these pockets, working in
the approximation that all observers form at the same time.

Compared to the previous discussion about late-time de Sitter
space nucleation, here $\Gamma^{\rm BB}_{{\rm reheat},ik}$ can be
much larger, since the temperature during reheating can be much
larger than $H_\Lambda$.  On the other hand, safety from
Boltzmann brains requires the late-time nucleation rate to be
small compared to the potentially very small vacuum decay rates,
while in this case the quantity on the right-hand side of
Eq.~(\ref{reheatbound}) is not exceptionally small.  In
discussing this issue, we will consider in sequence three
descriptions of the Boltzmann brain: a human-like brain, a
near-black hole computer, and a diffuse computer. 

The nucleation of human-like Boltzmann brains during reheating
was discussed in Ref.~\cite{LVW}, where it was pointed out that
such brains could not function at temperatures much higher than
300 K, and that the nucleation rate for a 100 kg object at this
temperature is \ $\sim \exp(-10^{40})$.  This suppression is
clearly more than enough to ensure that Eq.~(\ref{reheatbound})
is satisfied.

For a near-black hole computer with $I_{\rm BB} \approx S_{\rm
BB,max} \approx 10^{16}$, the minimum mass is 600 grams.  If we
assume that the reheat temperature is no more than the reduced
Planck mass, $m_{\rm Planck} \equiv 1/\sqrt{8 \pi G} \approx 2.4
\times 10^{18} \hbox{ GeV} \approx 4.3 \times 10^{-6}$ gram,
we find that $\Gamma^{\rm BB}_{\rm reheat} < \exp\left(-\sqrt{2
I_{\rm BB}} \right) \sim \exp(-10^8)$.  Although this is not
nearly as much suppression as in the previous case, it is clearly
enough to guarantee that Eq.~(\ref{reheatbound}) will be
satisfied. 

For the diffuse computer, we can consider an ideal gas of
massless particles, as discussed in
Eqs.~(\ref{pre-fourthirds})--(\ref{fourthirds}).  The system
would have approximately $S_{\rm max}$ particles, and a total
energy of $E = S_{\rm max}^{4/3}/R$, so the Boltzmann suppression
factor is $\exp\left[-S_{\rm max}^{4/3}/\left(R\, T_{\rm
reheat}\right)\right]$. The Boltzmann brain production can occur
at any time during the reheating process, so there is nothing
wrong with considering Boltzmann brain production in our universe
at the present time.  For $T_{\rm reheat} = 2.7$ K and $S_{\rm
max} = 10^{16}$, this formula implies that the exponent has
magnitude 1 for $R = S_{\rm max}^{4/3} T_{\rm reheat}^{-1}
\approx 200$ light-years.  Thus, the formula suggests that
diffuse-gas-cloud Boltzmann brains of radius 200 light-years can
be thermally produced in our universe, at the present time,
without suppression! If this estimate were valid, then Boltzmann
brains would almost certainly dominate the universe. 

We argue, however, that the gas clouds described above would have
no possibility of computing, because the thermal noise would
preclude any storage or transfer of information.  The entire
device has energy of order $E \approx T_{\rm reheat}$, which is
divided among approximately $10^{16}$ massless particles.  The
mean particle energy is therefore $10^{16}$ times smaller than
that of the thermal particles in the background radiation, and
the density of Boltzmann brain particles is $10^{48}$ times
smaller than the background.  To function, it seems reasonable
that the diffuse computer needs an energy per particle that is at
least comparable to the background, which means that the
suppression factor is $\exp(-10^{16})$ or smaller.  Thus, we
conclude that for all three cases, the ratio of Boltzmann brains
to normal observers is totally negligible.

Finally, let us also mention the possibility that Boltzmann
brains might form as quantum fluctuations in stable Minkowski
vacua.  String theory implies at least the existence of a 10D
decompactified Minkowski vacuum; Minkowski vacua of lower
dimension are not excluded, but they require precise fine tunings
for which motivation is lacking.  While quantum fluctuations in
Minkowski space are certainly less classical than in de Sitter
space, they still might be relevant. The possibility of Boltzmann
brains in Minkowski space has been suggested by Page
\cite{Page1,Page2,Page3}.  If $\Gamma_{\rm BB}$ is nonzero in
such vacua, regardless of how small it might be, Boltzmann brains
will always dominate in the scale-factor cutoff measure as we
have defined it. Even if Minkowski vacua cannot support Boltzmann
brains, there might still be a serious problem with what might be
called ``Boltzmann islands.'' That is, it is conceivable that a
fluctuation in a Minkowski vacuum can produce a small region of
an anthropic vacuum with a Boltzmann brain inside it.  The
anthropic vacuum could perhaps even have a different dimension
than its Minkowski parent.  If such a process has a nonvanishing
probability to occur, it will also give rise to Boltzmann brain
domination in the scale-factor cutoff measure. These problems
would be shared by all measures that assign an infinite weight to
stable Minkowski vacua.  There is, however, one further
complication which might allow Boltzmann brains to form in
Minkowski space without dominating the multiverse.  If one
speculates about Boltzmann brain production in Minkowski space,
one may equally well speculate about spontaneous creation of
inflationary universes there, each of which could contain
infinitely many normal observers \cite{Linde:1991sk}.  These
issues become complicated, and we will make no attempt to resolve
them here.  Fortunately, the estimates of thermal Boltzmann brain
nucleation rates in de Sitter space approach zero in the
Minkowski space limit $\Lambda\to 0$, so the issue of Boltzmann
brains formed by quantum fluctuations in Minkowski space can be
set aside for later study.  Hopefully the vague idea that these
fluctuations are less classical than de Sitter space fluctuations
can be promoted into a persuasive argument that they are not
relevant.


\subsection{Vacuum Decay Rates}

One of the most developed approaches to the string landscape
scenario is based on the KKLT construction~\cite{KKLT}.  In this
construction, one begins by finding a set of stabilized
supersymmetric AdS and Minkowski vacua.  After that, an uplifting
is performed, e.g. by adding a $\overline{D3}$ brane at the tip
of a conifold~\cite{KKLT}.  This uplifting makes the vacuum
energy density of some of these vacua positive (AdS $\to$ dS),
but in general many vacua remain AdS, and the Minkowski vacuum
corresponding to the uncompactified 10d space does not become
uplifted.  The enormous number of the vacua in the landscape
appears because of the large number of different topologies of
the compactified space, and the large number of different fluxes
and branes associated with it.

There are many ways in which our low-energy dS vacuum may decay.
First of all, it can always decay into the Minkowski vacuum
corresponding to the uncompactified 10d space~\cite{KKLT}. It can
also decay to one of the AdS vacua corresponding to the same set
of branes and fluxes~\cite{CDGKL}.  More generally, decays occur
due to the jumps between vacua with different fluxes, or due to
the brane-flux annihilation~\cite{Brown:1987dd,Brown:1988kg,
Bousso:2000xa,Kachru:2002gs,Frey:2003dm,Johnson:2008kc,
FreivogelLippert,BlancoPillado:2009di}, and may be accompanied by
a change in the number of compact
dimensions~\cite{Linde:1988yp,Carroll:2009dn,BlancoPillado:2009mi}.
If one does not take into account vacuum stabilization, these
transitions are relatively easy to
analyze~\cite{Brown:1987dd,Brown:1988kg,Bousso:2000xa}.  However,
in the realistic situations where the moduli fields are
determined by fluxes, branes, etc., these transitions involve a
simultaneous change of fluxes and various moduli fields, which
makes a detailed analysis of the tunneling quite complicated.

Therefore, we begin with an investigation of the simplest decay
modes due to the scalar field tunneling. The transition to the
10d Minkowski vacuum was analyzed in Ref.~\cite{KKLT}, where it
was shown that the decay rate $\kappa$ is always greater than
\beq
\kappa\,\gtrsim\, e^{-S_{D}} = 
\exp\!\left(-{24\pi^{2}\over V_{\rm dS}}\right)\,.
\label{KKLT1}
\eeq
Here $S_{D}$ is the entropy of dS space. For our vacuum, $S_{D}
\sim 10^{120}$, which yields
\beq
\kappa\,\gtrsim\, e^{-S_{D}}\sim \exp\left(-{10^{120}}\right)\,.
\label{KKLT2}
\eeq
Because of the inequality in Eq.~(\ref{KKLT1}), we expect the
slowest-decaying vacua to typically be those with very small
vacuum energies, with the dominant vacuum energy density possibly
being much smaller than the value in our universe.

The decay to AdS space (or, more accurately, a decay to a
collapsing open universe with a negative cosmological constant)
was studied in Ref.~\cite{CDGKL}.  The results of
Ref.~\cite{CDGKL} are based on investigation of BPS and near-BPS
domain walls in string theory, generalizing the results
previously obtained in $\N=1$ supergravity~\cite{Cvetic:1996vr,
Ceresole:2001wi,Behrndt:2001mx,Louis:2006wq}.  Here we briefly
summarize the main results obtained in Ref.~\cite{CDGKL}.

Consider, for simplicity, the situation where the tunneling
occurs between two vacua with very small vacuum energies.  For
the sake of argument, let us first ignore the gravitational
effects. Then the tunneling always takes place, as long as one
vacuum has higher vacuum energy than the other.  In the limit
when the difference between the vacuum energies goes to zero, the
radius of the bubble of the new vacuum becomes infinitely large,
$R \to \infty$ (the thin-wall limit). In this limit, the bubble
wall becomes flat, and its initial acceleration, at the moment
when the bubble forms, vanishes. Therefore to find the tension of
the domain wall in the thin wall approximation one should solve
an equation for the scalar field describing a static domain wall
separating the two vacua.

If the difference between the values of the scalar potential in
the two minima is too small, and at least one of them is AdS,
then the tunneling between them may be forbidden because of the
gravitational effects~\cite{cdl}.  In particular, all
supersymmetric vacua, including all KKLT vacua prior to the
uplifting, are absolutely stable even if other vacua with lower
energy density are available~\cite{deser,hullpositive,
wittenpositive,Weinberg:1982id}.

It is tempting to make a closely related but opposite statement:
non-supersymmetric vacua are always unstable.  However, this is
not always the case. In order to study tunneling while taking
account of supersymmetry (SUSY), one may start with two different
supersymmetric vacua in two different parts of the universe and
find a BPS domain wall separating them.  One can show that if the
superpotential does not change its sign on the way from one
vacuum to the other, then this domain wall plays the same role
as the flat domain wall in the no-gravity case discussed above:
it corresponds to the wall of the bubble that can be formed once
the supersymmetry is broken in either of the two minima. 
However, if the superpotential does change its sign, then only a
sufficiently large supersymmetry breaking will lead to the
tunneling~\cite{Cvetic:1996vr,CDGKL}.

One should keep this fact in mind, but since we are discussing a
landscape with an extremely large number of vacua, in what
follows we assume that there is at least one direction in which
the superpotential does not change its sign on the way from one
minimum to another.  In what follows we describe tunneling in one
such direction. Furthermore, we assume that at least some of the
AdS vacua to which our dS vacuum may decay are uplifted much less
than our vacuum.  This is a generic situation, since the
uplifting depends on the value of the volume modulus, which takes
different values in each vacuum.

In this case the decay rate of a dS vacuum with low energy
density and broken supersymmetry can be estimated as
follows~\cite{CDGKL,Dine:2007er}:
\beq
\kappa\sim\exp\!\left(-{8\pi^{2}\alpha\over 3 m_{3/2}^2}\right)\,,
\label{m3/2}
\eeq
where $m_{3/2}$ is the gravitino mass in that vacuum and $\alpha$
is a quantity that depends on the parameters of the potential. 
Generically one can expect $\alpha ={\mathcal O}(1)$, but it can
also be much greater or much smaller than ${\mathcal O}(1)$.  The
mass $m_{3/2}$ is set by the scale of SUSY breaking,
\beq
3m^{2}_{3/2}=\Lambda_{\rm SUSY}^4 \ ,
\eeq
where we recall that we use reduced Planck units, $8\pi G = 1$. 
Therefore the decay rate can be also represented in terms of the
SUSY-breaking scale $\Lambda_{\rm SUSY}$:
\beq
\kappa\sim\exp\!\left(-{24\pi^{2}\alpha\over \Lambda_{\rm SUSY}^4}\right)\,,
\label{m3/2a}
\eeq
Note that in the KKLT theory, $\Lambda_{\rm SUSY}^4$ corresponds
to the depth of the AdS vacuum before the uplifting, so that
\beq
\kappa\sim\exp\!\left(-{24\pi^{2}\alpha\over |V_{\rm AdS}|}\right)\,.
\label{m3/2b}
\eeq
In this form, the result for the tunneling looks very similar to
the lower bound on the decay rate of a dS vacuum,
Eq.~(\ref{KKLT1}), with the obvious replacements $\alpha \to 1$
and $|V_{\rm AdS}| \to V_{\rm dS}$.

Let us apply this result to the question of vacuum decay in our
universe.  Clearly, the implications of Eq.~(\ref{m3/2a}) depend
on the details of SUSY phenomenology. The standard requirement
that the gaugino mass and the scalar masses are ${\mathcal O}(1)$
TeV leads to the lower bound
\beq
\Lambda_{\rm SUSY} \gtrsim 10^{4}\mbox{--}10^{5}\,\,{\rm GeV} \ ,
\label{m3/2c}
\eeq
which can be reached, e.g., in the models of conformal gauge
mediation~\cite{Ibe:2008si}.  This implies that for our vacuum
\beq
\kappa_{\rm our}\gtrsim \exp(-10^{56})\mbox{--}\exp(-10^{60})\,.
\label{kour1}
\eeq
Using Eq.~(\ref{Gamma1}), the Boltzmann brain nucleation rate in
our universe exceeds the lower bound of the above inequality only
if $M\lesssim 10^{-9}$~kg.

On the other hand, one can imagine universes very similar to ours
except with much larger vacuum energy densities. The vacuum decay
rate of Eq.~(\ref{m3/2}) exceeds the Boltzmann brain nucleation
rate of Eq.~(\ref{Gamma1}) when
\beq
\bigg(\frac{m_{3/2}}{10^{-2}~{\rm eV}}\bigg)^{\!\!2}
\left(\frac{M}{1~{\rm kg}}\right)
\left(\frac{H_\Lambda^{-1}}{10^8~{\rm m}}\right) 
\gtrsim 10^9\alpha \,.
\label{comp}
\eeq
Note that $H_\Lambda^{-1}\sim 10^8$~m corresponds to the smallest
de Sitter radius for which the tidal force on a 10 cm brain does
not exceed the gravitational force on the surface of the earth,
while $m_{\rm 3/2}\sim 10^{-2}$~eV corresponds to $\Lambda_{\rm
SUSY}\sim 10^{4}$~GeV.  Thus, it appears the decay rate of
Eq.~(\ref{m3/2}) allows for Boltzmann brain domination. 

However, we do not really know whether the models with low
$\Lambda_{\rm SUSY}$ can successfully describe our world.  To
mention one potential problem: in models of string inflation
there is a generic constraint that during the last stage of
inflation one has $H \lesssim m_{{3/2}}$~\cite{Kallosh:2004yh}. 
If we assume the second and third factors of Eq.~(\ref{comp})
cannot be made much less than unity, then we only require
$m_{3/2} \gtrsim {\mathcal O}(10^{2})$~eV to avoid Boltzmann
brain domination.  While models of string inflation with
$H\lesssim 100$~eV are not entirely impossible in the string
landscape, they are extremely difficult to
construct~\cite{Conlon:2008cj}.  If instead of $\Lambda_{\rm
SUSY}\sim 10^{4}$~GeV one uses $\Lambda_{\rm SUSY}\sim
10^{11}$~GeV, as in models with gravity mediation, one finds
$m_{\rm 3/2}\sim 10^3$~GeV and Eq.~(\ref{comp}) is easily
satisfied.

These arguments apply when supersymmetry violation is as large or
larger than in our universe.  If supersymmetry violation is too
small, atomic systems are unstable~\cite{Susskindbook}, the
masses of some of the particles will change dramatically, etc. 
However, the Boltzmann computers described in the previous
subsection do not necessarily rely on laws of physics similar to
those in our universe (in fact, they seem to require very
different laws of physics).  The present authors are unaware of
an argument that supersymmetry breaking must be so strong that
vacuum decay is always faster than the Boltzmann brain production
rate of Eq. (\ref{computer}). 

On the other hand, up to this point we have used the estimates of
the vacuum decay rate that were obtained in
Refs.~\cite{CDGKL,Dine:2007er} by investigation of the transition
where only moduli fields changed.  As we have already mentioned,
the description of a more general class of transitions involving
the change of branes or fluxes is much more complicated. 
Investigation of such processes, performed in
Refs.~\cite{Kachru:2002gs,Frey:2003dm,FreivogelLippert},
indicates that the process of vacuum decay for any vacuum in the
KKLT scenario should be rather fast,
\beq
\kappa \gtrsim \exp(-10^{22})\,.
\label{kour3}
\eeq

The results of
Refs.~\cite{Kachru:2002gs,Frey:2003dm,FreivogelLippert}, like the
results of Refs.~\cite{CDGKL,Dine:2007er}, are not completely
generic.  In particular, the investigations of
Refs.~\cite{Kachru:2002gs,Frey:2003dm,FreivogelLippert} apply to
the original version of the KKLT scenario, where the uplifting of
the AdS vacuum occurs due to $\overline{D3}$ branes, but not to
its generalization proposed in Ref.~\cite{Burgess:2003ic}, where
the uplifting is achieved due to D7 branes.  Neither does it
apply to the recent version of dS stabilization proposed in
Ref.~\cite{Silverstein:2007ac}.  Nevertheless, the results of
Refs.~\cite{Kachru:2002gs,Frey:2003dm, FreivogelLippert} show
that the decay rate of dS vacua in the landscape can be quite
large. The rate $\kappa \gtrsim \exp(-10^{22})$ is much greater
than the expected rate of Boltzmann brain production given by
Eq.~(\ref{1051}).  However, it is just a bit smaller than the
bosonic gas Boltzmann brain production rate of Eq.~(\ref{1021})
and much smaller than our most dangerous upper bound on the
Boltzmann brain production rate, given by Eq.~(\ref{computer}).


\section{Conclusions}
\label{sec:conclusions}

If the observed accelerating expansion of the universe is driven
by constant vacuum energy density and if our universe does not
decay in the next 20 billion years or so, then it seems cosmology
must explain why we are ``normal observers'' --- who evolve from
non-equilibrium processes in the wake of the big bang --- as
opposed to ``Boltzmann brains'' --- freak observers that arise as
a result of rare quantum
fluctuations~\cite{Rees1,Albrecht,Susskind,Page06,BF06}.  Put in
experimental terms, cosmology must explain why we observe
structure formation in a residual cosmic microwave background, as
opposed to the empty, vacuum-energy dominated environment in
which almost all Boltzmann brains nucleate.  As vacuum-energy
expansion is eternal to the future, the number of Boltzmann
brains in an initially-finite comoving volume is infinite. 
However, if there exists a landscape of vacua, then rare
transitions to other vacua populate a diverging number of
universes in this comoving volume, creating an infinite number of
normal observers.  To weigh the relative number of Boltzmann
brains to normal observers requires a spacetime measure to
regulate the infinities.

Recently, the scale-factor cutoff measure was shown to possess a
number of desirable attributes, including avoiding the youngness
paradox~\cite{Tegmark:2004qd} and the $Q$ (and $G$)
catastrophe~\cite{FHW,QGV,GS}, while predicting the cosmological
constant to be measured in a range including the observed value,
and excluding values more than about a factor of ten larger and
smaller than this~\cite{DSGSV}.  The scale-factor cutoff does not
itself select for a longer duration of slow-roll inflation,
raising the possibility that a significant fraction of observers
like us measure cosmic curvature significantly above the value
expected from cosmic variance~\cite{DSGSV2}.  In this paper, we
have calculated the ratio of the total number of Boltzmann brains
to the number of normal observers, using the scale-factor cutoff.

The general conditions under which Boltzmann brain domination is
avoided were discussed in Subsection~\ref{ssec:genconds}, where
we described several alternative criteria that can be used to
ensure safety from Boltzmann brains.  We also explored a set of
assumptions that allow one to state conditions that are both
necessary and sufficient to avoid Boltzmann brain domination. 
One relatively simple way to ensure safety from Boltzmann brains
is to require two conditions: (1) in any vacuum, the Boltzmann
brain nucleation rate must be less than the decay rate of that
vacuum, and (2) for any anthropic vacuum $j$ with a decay rate
$\kappa_j
\approx q$, and for any non-anthropic vacuum $j$,
one must construct a sequence of transitions from $j$ to an
anthropic vacuum; if the sequence includes suppressed upward
jumps, then the Boltzmann brain nucleation rate in vacuum $j$
must be less than the decay rate of vacuum $j$ times the product
of all the suppressed branching ratios $B_{\rm up}$ that appear
in the sequence. The condition (2) might not be too difficult to
satisfy, since it will generically involve only states with very
low vacuum energy densities, which are likely to be nearly
supersymmetric and therefore unlikely to support the complex
structures needed for Boltzmann brains or normal observers. 
Condition (2) can also be satisfied if there is no unique
dominant vacuum, but instead a dominant vacuum system that
consists of a set of nearly degenerate states, some of which are
anthropic, which undergo rapid transitions to each other, but
only slow transitions to other states. The condition (1) is
perhaps more difficult to satisfy. Although nearly-supersymmetric
string vacua can in principle be
long-lived~\cite{Cvetic:1996vr,Ceresole:2001wi,Behrndt:2001mx,Louis:2006wq,
KKLT,CDGKL}, with decay rates possibly much smaller than the
Boltzmann brain nucleation rate, recent investigations suggest
that other decay channels may evade this
problem~\cite{Kachru:2002gs,Frey:2003dm,FreivogelLippert}.
However, the decay processes studied in
~\cite{Cvetic:1996vr,Ceresole:2001wi,Behrndt:2001mx,Louis:2006wq,
KKLT,CDGKL,Kachru:2002gs,Frey:2003dm,FreivogelLippert} do not
describe some of the situations which are possible in the string
theory landscape, and the strongest constraints on the decay rate
obtained in \cite{FreivogelLippert} are still insufficient to
guarantee that the vacuum decay rate is always smaller than the
fastest estimate of the Boltzmann brain production rate,
Eq.~(\ref{computer}).

One must emphasize that we are discussing a rapidly developing
field of knowledge. Our estimates of the Boltzmann brain
production rate are exponentially sensitive to our understanding
of what exactly the Boltzmann brain is. Similarly, the estimates
of the decay rate in the landscape became possible only five
years ago, and this subject certainly is going to evolve.
Therefore we will mention here two logical possibilities which
may emerge as a result of the further investigation of these
issues.

If further investigation will demonstrate that the Boltzmann
brain production rate is always smaller than the vacuum decay
rate in the landscape, the probability measure that we are
investigating in this paper will be shown not to suffer from the
Boltzmann brain problem. Conversely, if one believes that this
measure is correct, the fastest Boltzmann brain production rate
will give us a rather strong lower bound on the decay rate of the
metastable vacua in the landscape. We expect that similar
conclusions with respect to the Boltzmann brain problem should be
valid for the causal-patch measures~\cite{diamond,censor}.

On the other hand, if we do not find a sufficiently convincing
theoretical reason to believe that the vacuum decay rate in all
vacua in the landscape is always greater than the fastest
Boltzmann brain production rate, this would motivate the
consideration of other probability measures where the Boltzmann
brain problem can be solved even if the probability of their
production is not strongly suppressed.

In any case, our present understanding of the Boltzmann brain
problem does not rule out the scale-factor cutoff measure, but
the situation remains uncertain.

\begin{acknowledgments}
We thank Raphael Bousso, Ben Freivogel, I-Sheng Yang, Shamit
Kachru, Renata Kallosh, Delia Schwartz-Perlov, and Lenny Susskind
for useful discussion.  The work of ADS is supported in part by
the INFN ``Bruno Rossi'' Fellowship, and in part by the U.S. 
Department of Energy (DoE) under contract No. DE-FG02-05ER41360. 
AHG is supported in part by the DoE under contract No.
DE-FG02-05ER41360. AL and MN are supported by the NSF grant
0756174. MPS and AV are supported in part by the U.S. National
Science Foundation under grant NSF 322, and AV is also supported
in part by a grant from the Foundational Questions Institute
(FQXi).
\end{acknowledgments}

\appendix


\section{Boltzmann Brains in Schwarzschild--de Sitter Space}
\label{app:1}

As explained in Subsection~\ref{sec:BBnucleation},
Eq.~(\ref{Gamma2}) for the production rate of Boltzmann brains
must be reexamined when the Boltzmann brain radius becomes
comparable to the de Sitter radius. In this case we need to
describe the Boltzmann brain nucleation as a transition from an
initial state of empty de Sitter space with horizon radius
$H_\Lambda^{-1}$ to a final state in which the dS space is
altered by the presence of an object with mass $M$.  Assuming
that the object can be treated as spherically symmetric, the
space outside the object is described by the Schwarzschild--de
Sitter (SdS) metric \cite{Stuchlik:2008ea}:\footnote{We restore
$G=1/8\pi$ in this Appendix for clarity.}
\begin{eqnarray}
ds^2 &=& - \left( 1 - \frac{2GM}{r} - H_\Lambda^2 r^2 \right)
     dt^2 \nn \\
&& + \left( 1 - \frac{2GM}{r} - H_\Lambda^2 r^2
     \right)^{\!\!-1}\! dr^2 + r^2 \, d\Omega^2 \, . \,\,\,\,
\label{SdSmetric}
\end{eqnarray}
The SdS metric has two horizons, determined by the positive zeros
of $g_{tt}$, where the smaller and larger are called $R_{\rm
Sch}$ and $R_{\rm dS}$, respectively. We assume the Boltzmann
brain is stable but not a black hole, so its radius satisfies
$R_{\rm Sch}<R<R_{\rm dS}$.  The radii of the two horizons are
given by
\begin{eqnarray}
\begin{array}{rl}
 R_{\rm Sch}\!\! &= \displaystyle \frac{2}{\sqrt{3}\, H_\Lambda} 
\cos \left( \frac{\pi+\xi}{3} \right), \\
\noalign{\vskip 4pt}
 R_{\rm dS}\!\! &= \displaystyle \frac{2}{\sqrt{3}\, H_\Lambda} 
\cos \left( \frac{\pi-\xi}{3} \right),
\end{array}
\label{SdSsol}
\end{eqnarray}
where
\beq
\cos \xi = 3\sqrt{3}\, GMH_\Lambda\,.
\label{xidef}
\eeq
This last equation implies that for a given value of $H_\Lambda$,
there is an upper limit on how much mass can be contained within
the de Sitter horizon:
\beq
M \le M_{\rm max} = (3 \sqrt{3} G H_\Lambda)^{-1}\,.
\label{Hbound}
\eeq
Eqs.~(\ref{SdSsol}) and (\ref{xidef}) can be inverted to express
$M$ and $H_\Lambda$ in terms of the horizon radii:
\begin{eqnarray}
{1 \over H_\Lambda^2} &=& R_{\rm Sch}^2 + R_{\rm dS}^2 + R_{\rm
Sch} R_{\rm dS}
\label{inverseHLam} \\
M &=& {R_{\rm dS} \over 2 G} \left(1 - H_\Lambda^2 R_{\rm dS}^2
     \right) 
\label{inverseradii}\\
  &=& {R_{\rm Sch} \over 2 G} \left(1 - H_\Lambda^2 R_{\rm Sch}^2
     \right) \ .
\end{eqnarray}

We relate the Boltzmann brain nucleation rate to the decrease in
total entropy $\Delta S$ caused by the the nucleation process,
\beq
\Gamma_{\rm BB} \sim e^{- \Delta S} \, ,
\label{appDeltaS}
\eeq
where the final entropy is the sum of the entropies of the
Boltzmann brain and the de Sitter horizon. For a Boltzmann brain
with entropy $S_{\rm BB}$, the change in entropy is given by
\beq
\Delta S = \frac{\pi}{G}H_\Lambda^{-2} 
- \left( \frac{\pi}{G} R_{\rm dS}^2 + S_{\rm BB} \right) \, .
\label{DeltaS-SdS}
\eeq
Note that for small $M$ one can expand $\Delta S$ to find
\beq
\Delta S = {2\pi M \over H_\Lambda}-S_{\rm BB} + {\cal
     O}(G M^2)\,,
\label{smallM}
\eeq
giving a nucleation rate in agreement with
Eq.~(\ref{Gamma2}).\footnote{We thank Lenny Susskind for
explaining this method to us.}

\begin{figure}[t!]
\includegraphics[trim=5 0 0 0]{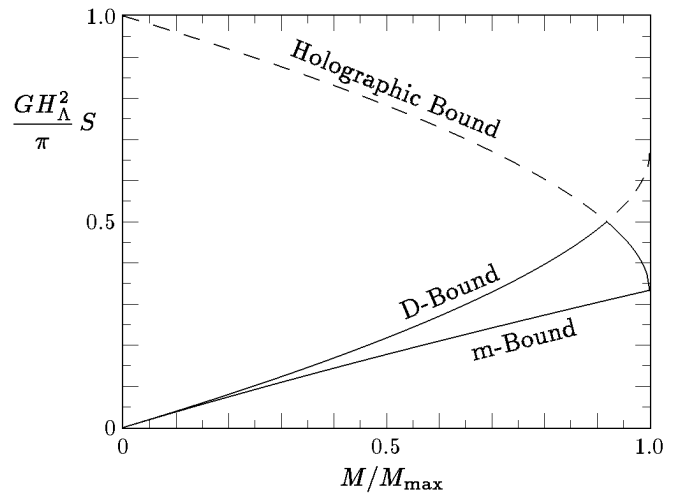}
\caption{Graph shows the holographic bound, the D-bound, and the
m-bound for the entropy of an object that fills de Sitter space
out to the horizon.  The holographic and D-- bounds are each
shown as broken lines in the region where they are superseded by
the other.  Although the m-bound looks very much like a straight
line, it is not.}
\label{fig:intbound}
\end{figure}

To find a bound on the nucleation rate, we need an upper bound on
the entropy that can be attained for a given size and mass.  In
flat space the entropy is believed to be bounded by Bekenstein's
formula, Eq.~(\ref{BekensteinB}), a bound which should also be
applicable whenever $R \ll R_{\rm dS}$.  More general bounds in
de Sitter space have been discussed by Bousso \cite{Bousso2000},
who considers bounds for systems that are allowed to fill the de
Sitter space out to the horizon $R=R_{\rm dS}$ of an observer
located at the origin.  For small mass $M$, Bousso argues that
the tightest known bound on $S$ is the D-bound, which states that
\beq
S \leq S_{\rm D} \equiv \frac{\pi}{G}
\left( \frac{1}{H_\Lambda^2} - R_{\rm dS}^2 \right) = 
\frac{\pi}{G}  \left( R_{\rm Sch}^2 + R_{\rm Sch} R_{\rm dS}
\right) \,,
\eeq
where the equality of the two expressions follows from
Eq.~(\ref{inverseHLam}).  This bound can be obtained from the
principle that the total entropy cannot increase when an object
disappears through the de Sitter horizon.  For larger values of
$M$, the tightest bound (for $R=R_{\rm dS}$) is the holographic
bound, which states that
\beq
S \leq S_{\rm H} \equiv \frac{\pi}{G} R_{\rm dS}^2\,.
\label{holographic}
\eeq
Bousso suggests the possibility that these bounds have a common
origin, in which case one would expect that there exists a valid
bound that interpolates smoothly between the two.  Specifically,
he points out that the function
\beq
S_{\rm m} \equiv \frac{\pi}{G} R_{\rm Sch} R_{\rm dS}\,
\label{interp}
\eeq
is a candidate for such a function.  Fig.~(\ref{fig:intbound})
shows a graph of the holographic bound, the D-bound, and the
m-bound (Eq.~(\ref{interp})) as a function of $M/M_{\rm max}$. 
While there is no reason to assume that $S_{\rm m}$ is a rigorous
bound, it is known to be valid in the extreme cases where it
reduces to the $D$-- and holographic bounds.  In between it might
be valid, but in any case it can be expected to be valid up to a
correction of order one.  In fact, Fig.~(\ref{fig:intbound}) and
the associated equations show that the worst possible violation
of the m-bound is at the point where the holographic and D--
bounds cross, at $M/M_{\rm max} = 3
\sqrt{6}/8 = 0.9186$, where the entropy can be no more than $(1 +
\sqrt{5})/2 = 1.6180$ times as large as $S_{\rm m}$. 

Here we wish to carry the notion of interpolation one step
further, because we would like to discuss in the same formalism
systems for which $R \ll R_{\rm dS}$, where the Bekenstein bound
should apply.  Hence we will explore the consequences of the
bound
\beq
S \leq S_{\rm I} \equiv \frac{\pi}{G} R_{\rm Sch} R \,,
 \label{guthbound}
\eeq
which we will call the interpolating bound.  This bound agrees
exactly with the m-bound when the object is allowed to fill de
Sitter space, with $R=R_{\rm dS}$.  Again we have no grounds to
assume that the bound is rigorously true, but we do know that it
is true in the three limiting cases where it reduces to the
Bekenstein bound, the $D$-bound, and the holographic bound.  The
limiting cases are generally the most interesting for us in any
case, since we wish to explore the limiting cases for Boltzmann
brain nucleation.  For parameters in between the limiting cases,
it again seems reasonable to assume that the bound is at least a
valid estimate, presumably accurate up to a factor of order one. 
We know of no rigorous entropy bounds for de Sitter space with
$R$ comparable to $R_{\rm dS}$ but not equal to it, so we don't
see any way at this time to do better than the interpolating
bound.

Proceeding with the I-bound of Eq.~(\ref{guthbound}), we can use
Eq.~(\ref{Idefine}) to rewrite Eq.~(\ref{DeltaS-SdS}) as
\beq
\Delta S = \frac{\pi}{G} \left( H_\Lambda^{-2} 
- R_{\rm dS}^2 \right) - S_{\rm BB, max} + I_{\rm BB} \, ,
\eeq
which can be combined with $S_{\rm BB, max} \le S_I$ to give
\beq
\Delta S \ge \frac{\pi}{G} \left( H_\Lambda^{-2} 
- R_{\rm dS}^2 - R_{\rm Sch}\,R\right) + I_{\rm BB} \, ,
\eeq
which can then be simplified using Eq.~(\ref{inverseHLam}) to
give
\beq
\Delta S \ge \frac{\pi}{G} R_{\rm Sch} \left( R_{\rm Sch} +
R_{\rm dS} - R\right) + I_{\rm BB} \, .
\label{Lastbound}
\eeq

\begin{figure}[t!]
\includegraphics[trim=5 0 0 0]{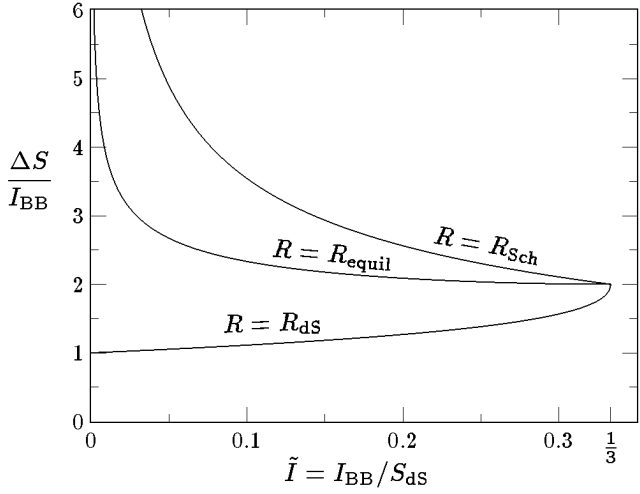}
\caption{Graph shows the ratio of $\Delta S$ to $I_{\rm BB}$,
where the nucleation rate for Boltzmann brains is proportional to
$e^{- \Delta S}$.  All curves are based on the $I$-bound, as
discussed in the text, but they differ by their assumptions about
the size $R$ of the Boltzmann brain.}
\label{fig:deltas}
\end{figure}

To continue, we have to decide what possibilities to consider for
the radius $R$ of the Boltzmann brain, which is related to the
question of Boltzmann brain stabilization discussed after
Eq.~(\ref{BBlimit}).  If we assume that stabilization is not a
problem, because it can be achieved by a domain wall or by some
other particle physics mechanism, then $\Delta S$ is minimized by
taking $R$ at its maximum value, $R = R_{\rm dS}$, so
\beq
\Delta S \ge \frac{\pi}{G} R_{\rm Sch}^2 + I_{\rm BB} \, .
\label{BoundRds}
\eeq
$\Delta S$ is then minimized by taking the minimum possible value
of $R_{\rm Sch}$, which is the value that is just large enough to
allow the required entropy, $S_{\rm BB,max} \ge I_{\rm BB}$. 
Using again the I-bound, one finds that saturation of the bound
occurs at
\beq
\xi_{\rm sat} = 3 \sin^{-1}\left( {{\sqrt{1 - 3 \tilde I} \over 2}
     }\right) \ ,
\eeq
where
\beq
\tilde I \equiv {I_{\rm BB} \over S_{\rm dS}} = {G H_\Lambda^2
\over \pi} I_{\rm BB} 
\eeq
is the ratio of the Boltzmann brain information to the entropy of
the unperturbed de Sitter space.  Note that $\tilde I$ varies
from zero to a maximum value of 1/3, which occurs in the limiting
case for which $R_{\rm Sch} = R_{\rm dS}$.  The saturating value
of the mass and the corresponding values of the Schwarzschild
radius and de Sitter radius are given by
\begin{eqnarray}
M_{\rm sat} &=& {\tilde I \sqrt{1 + \tilde I} \over 2 G
H_\Lambda} \ ,\\
R_{\rm Sch,sat} &=& {\sqrt{1+\tilde I} - \sqrt{1 - 3
\tilde I} \over 2 H_\Lambda} \, ,\\
R_{\rm dS,sat} &=& {\sqrt{1 - 3 \tilde I} + \sqrt{1 + \tilde I}
\over 2 H_\Lambda} \ .
\end{eqnarray}
Combining these results with Eq.~(\ref{BoundRds}), one has for
this case ($R=R_{\rm dS}$) the bound
\beq
{\Delta S \over I_{\rm BB}} \ge {1 + \tilde I - \sqrt{1 + \tilde
I} \, \sqrt{1 - 3 \tilde I} \over 2 \tilde I} \, .
\label{Bound1}
\eeq
As can be seen in Figure \ref{fig:deltas}, the bound on $\Delta
S/I_{\rm BB}$ for this case varies from 1, in the limit of
vanishing $\tilde I$ (or equivalently, the limit $H_\Lambda
\to 0$), to 2, in the limit $R_{\rm Sch} \to R_{\rm dS}$.

The limiting case of $\tilde I_{\rm BB} \to 0$, with a nucleation
rate of order $e^{-I_{\rm BB}}$, has some peculiar features that
are worth mentioning.  The nucleation rate describes the
nucleation of a Boltzmann brain with some particular memory
state, so there would be an extra factor of $e^{I_{\rm BB}}$ in
the sum over all memory states.  Thus, a single-state nucleation
rate of $e^{-I_{\rm BB}}$ indicates that the total nucleation
rate, including all memory states, is not suppressed at all.  It
may seem strange that the nucleation rate could be unsuppressed,
but one must keep in mind that the system will function as a
Boltzmann brain only for very special values of the memory state. 
In the limiting case discussed here, the ``Boltzmann brain''
takes the form of a minor perturbation of the degrees of freedom
associated with the de Sitter entropy $S_{\rm dS} = \pi / (G
H_\Lambda^2) $.

As a second possibility for the radius $R$, we can consider the
case of strong gravitational binding, $R \to R_{\rm Sch}$, as
discussed following Eq.~(\ref{BBlimit}).  For this case the bound
(\ref{Lastbound}) becomes
\beq
\Delta S \ge \frac{\pi}{G} R_{\rm Sch} R_{\rm dS} + I_{\rm BB} \, .
\label{BoundRSch}
\eeq
(Interestingly, if we take $I=0$ ($S_{\rm BB} = S_{\rm max}$)
this formula agrees with the result found in
Ref.~\cite{Bousso-Hawking} for black hole nucleation in de Sitter
space.) With $R = R_{\rm Sch}$ the saturation of the I-bound
occurs at
\beq
\xi_{\rm sat} = {\pi \over 2} - 3 \sin^{-1} \left(\sqrt{ 3 \tilde
I} \over 2\right) \ .
\eeq
The saturating value of the mass and the corresponding values of
the Schwarzschild radius and de Sitter radius are given by
\begin{eqnarray}
M_{\rm sat} &=& {\sqrt{\tilde I} \, \left(1 - \tilde I \right) \over 2 G
H_\Lambda} \ ,\\
R_{\rm Sch,sat} &=& {\sqrt{ \tilde I} \over H_\Lambda} \, ,\\
R_{\rm dS,sat} &=& {\sqrt{4 - 3 \tilde I} - \sqrt{\tilde I} \over
2 H_\Lambda}  \ .
\end{eqnarray}
Using these relations to evaluate $\Delta S$ from
Eq.~(\ref{BoundRSch}), one finds
\beq
{\Delta S \over I_{\rm BB}} = {\sqrt{4 - 3 \tilde I} +
     \sqrt{\tilde I} \over 2 \sqrt{\tilde I}} \ ,
\eeq
which is also plotted in Figure \ref{fig:deltas}.  In this case
($R=R_{\rm Sch}$) the smallest ratio $\Delta S/I_{\rm BB}$ is 2,
occurring at $\tilde I = 1/3$, where $R_{\rm Sch} = R_{\rm dS}$. 
For smaller values of $\tilde I$ the ratio becomes larger,
blowing up as $1/\sqrt{\tilde I}$ for small $\tilde I$.  Thus,
the nucleation rates for this choice of $R$ will be considerably
smaller than those for Boltzmann brains with $R \approx R_{\rm
dS}$, but this case would still be relevant in cases where
Boltzmann brains with $R \approx R_{\rm dS}$ cannot be
stabilized.

Another interesting case, which we will consider, is to allow the
Boltzmann brain to extend to $R = R_{\rm equil}$, the point of
equilibrium between the gravitational attraction of the Boltzmann
brain and the outward gravitational pull of the de Sitter
expansion.  This equilibrium occurs at the stationary point of
$g_{tt}$, which gives
\beq
R_{\rm equil} = \left( {G M \over H_\Lambda^2 } \right)^{1/3} \,
.
\eeq
Boltzmann brains within this radius bound would not be pulled by
the de Sitter expansion, so relatively small mechanical forces
will be sufficient to hold them together. 

Again $\Delta S$ will be minimized when the $I$-bound is
saturated, which in this case occurs when
\beq
\xi_{\rm sat} = {\pi \over 2} - 3 \sin^{-1} \left[ {\sqrt{1 - 2
A(\tilde I) } \over 2} \right] \, ,
\eeq
where
\beq
A(\tilde I) \equiv \sin \left[ {\sin^{-1} \left(1 - 27 \,
\tilde I\,^{\raise 2pt\hbox{$\scriptstyle 3$}}
\right) \over 3} \right] \, .
\eeq
The saturating value of the mass and the Schwarzschild and de
Sitter radii are given by
\begin{eqnarray}
M_{\rm sat} &=& {\sqrt{3} [1+A(\tilde I) ] \sqrt{1 - 2 A(\tilde
I) } \over 9 G H_\Lambda} \, ,\\
R_{\rm Sch, sat} &=& {\sqrt{1 - 2 A(\tilde I)} \over \sqrt{3}\, H_\Lambda
} \, ,\\
R_{\rm dS, sat} &=& { \sqrt{3} \left[ \sqrt{3 \left(3 + 2A(\tilde
I) \right)} - \sqrt{1 - 2 A(\tilde I)} \right] \over 6 H_\Lambda}
\, .\nn\\
\noalign{\vskip-8pt}
\end{eqnarray}
The equilibrium radius itself is given by
\beq
R_{\rm equil, sat} = {\left[ 1 - 2 A(\tilde I)\right]^{1/6}
\left[1 + A(\tilde I) \right]^{1/3} \over \sqrt{3} H_\Lambda} \,
.
\eeq
Using these results with Eq.~(\ref{Lastbound}), $\Delta S$ is
found to be bounded by
\beq {\Delta S \over I_{\rm BB}} = {\sqrt{3 \left(1 - 2
A(\tilde I) \right) \left( 3 + 2 A(\tilde I)\right)} - 2 A(\tilde
I) + 1 \over 6 \tilde I } \, ,
\eeq
which is also plotted in Figure \ref{fig:deltas}.  As one might
expect it is intermediate between the two other cases.  Like the
$R=R_{\rm Sch}$ case, however, the ratio $\Delta S/I_{\rm BB}$
blows up for small $\tilde I$, in this case behaving as
$(2/\tilde I)^{1/4}$.

In summary, we have found that our study of tunneling in
Schwarzschild--de Sitter space confirms the qualitative
conclusions that were described in
Subsection~\ref{sec:BBnucleation}.  In particular, we have found
that if the entropy bound can be saturated, then the nucleation
rate of a Boltzmann brain requiring information content $I_{\rm
BB}$ is given approximately by $e^{- a I_{\rm BB}}$, where $a$ is
of order a few, as in Eq.~(\ref{BBlimit}).  The coefficient $a$
is always greater than 2 for Boltzmann brains that are small
enough to be gravitationally bound.  This conclusion applies
whether one insists that they be near-black holes, or whether one
merely requires that they be small enough so that their
self-gravity overcomes the de Sitter expansion.  If, however, one
considers Boltzmann brains whose radius is allowed to extend to
the de Sitter horizon, then Figure \ref{fig:deltas} shows that
$a$ can come arbitrarily close to 1.  However, one must remember
that the $R=R_{\rm dS}$ curve on Figure \ref{fig:deltas} can be
reached only if several barriers can be overcome.  First, these
objects are large and diffuse, becoming more and more diffuse as
$\tilde I$ approaches zero and $a$ approaches 1.  There is no
known way to saturate the entropy bound for such diffuse systems,
and Eq.~(\ref{fourthirds}) shows that an ideal gas model leads to
$a \sim I_{\rm BB}^{1/3} \gg 1$.  Furthermore, Boltzmann brains
of this size can function only if some particle physics mechanism
is available to stabilize them against the de Sitter expansion. 
A domain wall provides a simple example of such a mechanism, but
Eq.~(\ref{ShellConstraint2}) indicates that the domain wall
solution is an option only if a domain wall exists with tension
$\sigma \sim I_{\rm BB} H_\Lambda^3$.  Thus, it is not clear how
close $a$ can come to its limiting value of 1.  Finally, we
should keep in mind that it is not clear if any of the examples
discussed in this appendix can actually be attained, since black
holes might be the only objects that saturate the entropy bound
for $S \gg 1.$

\end{document}